%% file: PHY-AE.tex
\documentclass[journal, final]{IEEEtran}
\usepackage[nolist]{acronym}
\usepackage{tikz}
\usetikzlibrary{decorations.pathreplacing,calligraphy}
\usepackage{pgfplots}
\usepgfplotslibrary{groupplots}
\usepackage{graphicx}
\usepackage{subfloat}
\usepackage{caption}
\usepackage{subcaption}
\usepackage{float}
\pgfplotsset{compat=newest}
\usepackage{units}
\usepackage{amsmath, amsbsy, amssymb}
\usepackage{tikzscale}
\usepackage{color}
\usepackage{epigraph}
\usepackage{circuitikz}
\usepackage{multirow}
\usepackage{booktabs}
\usepackage[ruled]{algorithm2e}

\usepackage[group-separator={,}]{siunitx}
\definecolor{darkgreen}{rgb}{0.125,0.5,0.169}
\usetikzlibrary{shapes,arrows,fadings}
\usepackage{marvosym}
\usepackage{bbm}
\usepackage{mathtools}

\tikzset{>=latex}

\input{macros.tex}

\input{acronyms.tex}
\input{corporateColours.tex}

\input{plotcyclelists.tex}

\usetikzlibrary{external}
\IEEEoverridecommandlockouts

\begin{document}

\title{Learning Joint Detection, Equalization and Decoding for Short-Packet Communications}

\author{\IEEEauthorblockN{Sebastian D\"orner$^{1}$, Jannis Clausius$^{1}$, Sebastian Cammerer$^{2}$, and Stephan ten Brink$^{1}$}

\IEEEauthorblockA{
$^{1}$ Institute of Telecommunications, University of Stuttgart, Pfaffenwaldring 47, 70659 Stuttgart, Germany \\
\{doerner,clausius,tenbrink\}@inue.uni-stuttgart.de\\
$^{2}$ NVIDIA, Fasanenstra{\ss}e 81, 10623 Berlin, Germany \\
scammerer@nvidia.com
}

\thanks{This work is supported by the German Federal Ministry of Education and Research (BMBF) within the project Open6GHub under grant 16KISK019 and the project FunKI  under grant 16KIS1187.}
}

\maketitle

\begin{abstract}
We propose and practically demonstrate a joint detection and decoding scheme for short-packet wireless communications in scenarios that require to first detect the presence of a message before actually decoding it.
For this, we extend the recently proposed serial Turbo-autoencoder \ac{NN} architecture and train it to find short messages that can be, all ``at once'', detected, synchronized, equalized and decoded when sent over an unsynchronized channel with memory.
The conceptional advantage of the proposed system stems from a holistic message structure with superimposed pilots for joint detection and decoding without the need of relying on a dedicated preamble.
This results not only in higher spectral efficiency, but also translates into the possibility of shorter messages compared to using a dedicated preamble.
We compare the \ac{DER}, \ac{BER} and \ac{BLER} performance of the proposed system with a hand-crafted state-of-the-art conventional baseline and our simulations show a significant advantage of the proposed autoencoder-based system over the conventional baseline in every scenario up to messages conveying $k\!=\!96$ information bits.
Finally, we practically evaluate and confirm the improved performance of the proposed system \ac{OTA} using a \ac{SDR}-based measurement testbed.
\end{abstract}

\acresetall

\section{Introduction}

In the last decades, wireless communication systems have been continuously pushed towards their fundamental physical limits and roughly 70 years after Shannon has quantified the achievable performance bounds in his seminal work \cite{shannon48theory}, we virtually achieve his capacity limits over many channels \cite{chung2001design}. %
However, increased connectivity and the expected omnipresence of smart devices in almost all fields of our daily life becomes a challenging -- but yet exciting -- requirement for future communication systems and upcoming wireless communication standards \cite{5GTechCommMag,You2020}. %
From a wireless engineering perspective, these \ac{IoT} and \ac{mMTC} networks are mostly characterized by their large number of devices. %
It is also expected that many of those low-rate devices do not require the transmission of large amounts of data, but short messages with sporadic channel accesses \cite{5GTechCommMag}.

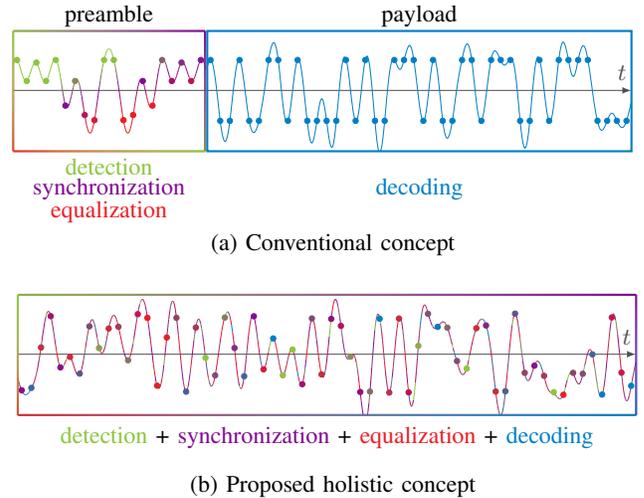
\begin{figure}[t]
\centering
\begin{subfigure}[b]{\columnwidth}
    \centering
    \input{tikz/bl_message_5.tikz}
    \vspace{-0.2cm}
    \caption{Conventional concept}
    \label{fig:baseline_blurb}
\end{subfigure}\\[3ex]
\begin{subfigure}[b]{\columnwidth}
    \centering
    \input{tikz/phyae_message_7.tikz}
    \caption{Proposed holistic concept}
\end{subfigure}
\caption{Conventional system with a dedicated preamble structure compared to the proposed holistically learned message.}
\label{fig:basic_blurb}
\end{figure}

Yet, when considering the actual end-to-end performance of today's systems for short packet transmissions, we are far away from Shannon's achievable rates as the evaluation of systems only based on their symbol-wise spectral efficiency over well-defined and synchronized channels does often not constitute their practical performance. %
Especially for short messages where a conventional dedicated preamble constitutes severe overhead, a carefully designed procedure is required for detection, channel estimation and to synchronize transmitter and receiver.
However, from an information theoretical perspective it is known that the achievable coding rate diminishes for short block lengths \cite{Polyanskiy2010} and the conventional separation of a transmission into a dedicated pilot preamble for detection/estimation and a payload sequence carrying the information, as depicted in Fig.~\ref{fig:basic_blurb}, is provably suboptimal \cite{lancho21jointdetdec,8461650}.
The theoretical results in \cite{lancho21jointdetdec} for the binary-input \ac{AWGN} channel further indicate, that ``in the short-packet regime, joint detection and decoding yields significant gains in terms of maximum coding rate over preamble-based detection followed by decoding'' \cite[p.~5]{lancho21jointdetdec}.

\begin{figure}[t]
	\centering
	\resizebox{0.475\textwidth}{!}{\input{tikz/system_model}}
	\caption{System overview showing all considered tasks in the scope of this work including channel coding, modulation, detection, synchronization, equalization and decoding. %
	}
	\label{fig:system_model_tikz}
	\vspace*{-0.4cm}
\end{figure}
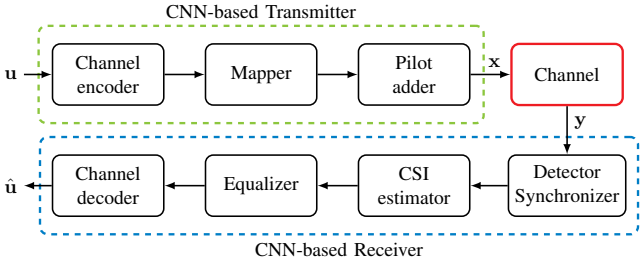

In this work, we propose to jointly learn the transmitter and receiver end-to-end for the aforementioned tasks of detection, synchronization, equalization and decoding (see Fig.~\ref{fig:basic_blurb}) by following the concept of autoencoder-based communications \cite{o2017introduction}.
The main contributions of this paper are: %
\begin{itemize}
\item\ We demonstrate a fully differentiable end-to-end learning scheme to establish the whole communications link on the \ac{PHY} layer using ultra short holistic messages.
The proposed system is based on an enhanced serial Turbo-autoencoder \cite{clausius21serialAE} and the task goes even beyond joint detection and decoding \cite{lancho21jointdetdec} as our system model, depicted in Fig.~\ref{fig:system_model_tikz}, also includes synchronization and equalization over multipath channels.%
\item As we implicitly consider detection, synchronization, equalization and decoding in the applied end-to-end performance metric, the transmitter is trained to find an optimal trade-off between detectability and the amount of added redundancy for decoding of the information payload.
For further insights and an intuitive interpretation of the learned solution, we investigate and evaluate the messages learned on the transmitter side.
\item We benchmark the performance of the proposed system against a conventional state-of-the-art baseline and show significant gains for all simulated scenarios up to messages conveying $k\!=\!96$ information bits.
\item Finally, we demonstrate the practicability of the proposed concept over-the-air within a \ac{SDR}-based measurement testbed and discuss the implications of actual deployment.
\end{itemize}

The rest of this paper is organized as follows.
Section~\ref{sec:related} provides a short overview of related literature.
Section~\ref{sec:sys} introduces the system model, the channel model and the corresponding conventional baseline system.
The proposed holistic end-to-end framework is introduced in Section~\ref{sec:ae}.
In Section~\ref{sec:results} we present the results of our simulations and over-the-air measurements and Section~\ref{sec:conclusion} concludes the paper.

\paragraph*{Notations}
\textcolor{black}{
Random variables are denoted by capital italic font, e.g., $X ,Y$, with realizations $x, y$, respectively.
Vectors are represented using a lower case bold font, e.g., $\mathbf{y}$.
$\mathcal{U}(a,b)$ denotes a uniform distribution in the range $[a,b)$ and
$\mathcal{N}(\mu,\sigma^2)$ denotes a normal distribution with mean $\mu$ and variance $\sigma^2$.
}

\section{Related Work}
\label{sec:related}

Autoencoder-based communication \cite{o2017introduction} has provided a new paradigm of how to design future communication systems -- instead of individually optimized signal-processing blocks, we can now jointly optimize the whole transceiver w.r.t. a single end-to-end performance metric.
The first synchronization for end-to-end learning has already been proposed in \cite{Doerner2018}, however, the performance was limited due to the very short messages of only a few channel uses and the synchronization \ac{NN} was trained consecutively after the messages were learned, thus, no strict joint optimization.
For most of the follow-up work \cite{FaycalTrimming2021,jiang19Turboae,Karanov:18,GunduzSourceChannelCoding2019,cammerer2019tcom}, the focus so far has mostly been on the advantages of joint signal processing and the possibilities of data-driven system optimization \cite{Aoudia2019,cammerer2019tcom,doerner2020gan}, i.e., the system can be trained with real-world data including all actual hardware and channel impairments.
Also, with the emergence of the Turbo-autoencoder architecture \cite{jiang19Turboae}, which allows an increased amount of information bits carried within a message, the focus was mostly on learning a coded modulation scheme for the \ac{AWGN} channel that can compete with state-of-the-art short length channel codes or even outperform state-of-the-art for channels with feedback \cite{jiang2020feedbackturboae}.
This concept has been extended to serially concatenated Turbo-autoencoders in \cite{clausius21serialAE} for \ac{AWGN} channels.
Using the autoencoder scheme for channels with inter-symbol-interference was investigated in \cite{zhang21autoencoder_isi}, however for a static channel model without detection or synchronization.
It is worth mentioning that the concept of hyper-dimensional modulation \cite{KimHDM2018} describes a similar, but non-differentiable, idea of learning holistic messages, which could be potentially also extended by an outer synchronization component. %
Also the authors of \cite{BianBIGAMP2021} identified a similar \ac{mMTC} scenario and addressed the task of joint detection, channel estimation and decoding of conventional messages with a Turbo-receiver scheme using a bilinear generalized approximate message passing (BiG-AMP) algorithm.
In this work, we want to shed light on the question of how to materialize the full end-to-end learning potential of the autoencoder concept by not just focusing on joint signal-processing capabilities and raw coding performance, but also on the advantages of learning a holistic message structure that is inherently introduced by the Turbo-autoencoder architecture as visualized in Fig.~\ref{fig:basic_blurb}.

\section{System Model}
\label{sec:sys}

We assume a \ac{SISO} system model with single-carrier modulation as depicted in Fig.~\ref{fig:system_model_tikz}.
The objective is to transmit $k$ bits within a short-packet message of $n$ complex-valued channel uses (symbols) including all required overhead for synchronization.
In this scenario, the channel model adds a random offset $\tau_\text{off} \in (0,n)$ to the message starting point to model initial random access without any prior signaling.
In conventional systems this is commonly done using a message that consists of a dedicated preamble sequence followed by the individual payload.
Typically, the payload modulation consists of symbols chosen from a set of alphabets, e.g., \ac{QPSK}, 16-\ac{QAM} or 256-\ac{QAM} which, however, may be a user specific choice, e.g., depending on the individual SNR.
The problem can be further complicated by additional \ac{CFO} and \ac{SFO} as transmitter and receiver oscillators do not share the exact same frequency.
Typically, a preamble sequence is used for detection and to establish timing synchronization and only if this is ensured, the processing of the remaining data is possible.
One example of such preamble sequences are Zadoff-Chu sequences, which are also used in the 3GPP 5G standard \cite{3gpp5G_phychannels}.
The preamble sequence can then also be used for channel estimation to enable equalization and decoding of the following payload data.
Additional pilots may be embedded in the message structure to further improve this process.
However, in particular for systems that only aim to transmit a short-packet message (i.e., few information bits, e.g. $k < 100$), the signaling overhead to establish timing synchronization between transmitter and receiver and for piloting of the channel can be large.

Given this general description, we want to quickly clarify the terminology throughout this work by defining the meaning\footnote{Note that some terms are not clearly defined in literature and their exact meaning often depends on the application. However, these are the definitions we use throughout this work.} of the following terms: %
\begin{itemize}
\item \textbf{Message:} A message $\xv$ consists of $n$ complex-valued channel uses (symbols), carrying $k$ information bits in total. This results in a message's information rate $R = \nicefrac{k}{n}$ (not to be confused with the code rate $R_c = \nicefrac{k}{k_\text{coded}}$ of conventional systems using a coded payload).
\item \textbf{Detection:} The receiver scans received sequences $\yv$ of length $n_\text{det} = 2 \cdot n$ and must be able to detect whether a full message lies within such a received sequence of complex-valued symbols or not.
\item \textbf{Synchronization:} The receiver needs to synchronize if a message has been correctly detected. This means the receiver must estimate the exact time offset $\hat{\tau}_\text{off}$ of the start/end of the message within the received sequence $\yv$.
\item \textbf{Decoding:} Recovery of the originally transmitted information bits. We define the reliability of the system by the \ac{BER} and \ac{BLER} performance after successful detection and synchronization.
\end{itemize}

\subsection{Channel Model}
\label{sec:channel_model}

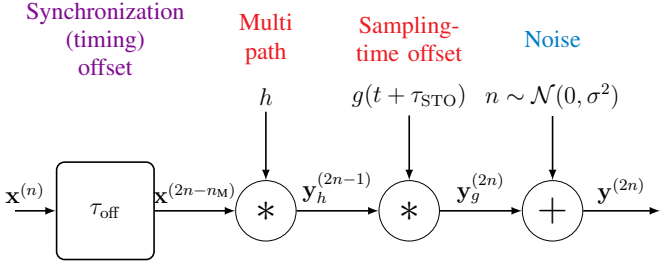
\begin{figure}[t]
	\centering
	\resizebox{0.5\textwidth}{!}{\input{tikz/channel_model_tikz}}
	\caption{Channel model used for training and performance evaluation.}

	\label{fig:channel_model_tikz}
	\vspace*{-0.4cm}
\end{figure}

For performance simulations and training we employ a stochastic multi-path propagation channel model derived from a Proakis \emph{type C} tap-delay-line model with five coefficients \cite{proakis2001digital}.
The $n_\text{taps}=5$ channel taps $h_i$ are randomly drawn for each channel realization, following a normal-distribution with Proakis weighting coefficients according to $h_i = w_{\text{Proakis},i} \cdot t_i$, where $t_i \sim \mathcal{N}(0,1)$ and $\wv_\text{Proakis} = [0.227, 0.46, 0.688, 0.46, 0.227]$.
The full channel model is depicted in Fig.~\ref{fig:channel_model_tikz}.
In a first step the random channel access of a device is modeled by adding a random delay of $\tau_\text{off} \sim \mathcal{U}(0,n-n_\text{M})$, where $n_\text{M}$ is the memory length of the channel, to the transmitted message $\xv$.
The resulting output $\xv^{(2n-n_\text{M})}$ is a sequence of $2n-n_\text{M}$ symbols consisting of the message $\xv$ with random delay according to $\tau_\text{off}$ and zero padding.
This sequence is then convolved with a randomly drawn channel impulse response $\hv$, resulting in $\yv_h^{(2n-1)}$ of length $2n-1$ due to a full convolution.
To simulate a random \ac{STO} at the receiver, as the result of un-synchronized oscillators between transmitter and receiver, $\yv_h^{(2n-1)}$ is then convolved with a randomly phase-shifted raised-cosine-filter vector $\gv_\text{STO}^{(16)}$ sampled from $g(t+\tau_\text{STO})$ with
\begin{equation}
	g(t) =
	\begin{cases}
		1 & \text{$t$=0}\\
		\frac{\pi}{4}\operatorname{sinc}\left(\frac{\pi}{2 \beta}\right) & \text{$|t|=\frac{T}{2\beta}$}\\
		\frac{\cos \left(\beta t /T\right)}{\pi t /T}\operatorname{sinc}\left(\frac{\pi t}{T}\right) & \text{otherwise.}
	\end{cases}
\end{equation}
Note that we assume, that the \ac{STO} adds a single significant tap to the impulse response which results in a total memory of the channel of $n_\text{M}=n_\text{taps}-1+1$ and the signal $\yv_g^{(2n)}$ of length $2n$.
Finally, we add white Gaussian noise $\nv \sim \mathcal{N}(0,\sigma^2)$ to $\yv_g^{(2n)}$ resulting in $\yv^{(2n)}$ as our channel model's output sequence.
We decided not to model \ac{CFO} and \ac{SFO} as the influence of \ac{SFO} is negligible for ultra short sequences and the compensation of notable \ac{CFO} would pose a significant challenge for the conventional baseline system since it would probably require additional pilot symbols at the end of a message.
Also, considering the ``not clock-synced'' \ac{SDR}-based testbed, which was used to evaluate actual deployment, the influence of frequency offsets was existent but negligible throughout our measurements. %

\subsection{Conventional State-of-the-Art Baseline System}
\label{sec:baseline}

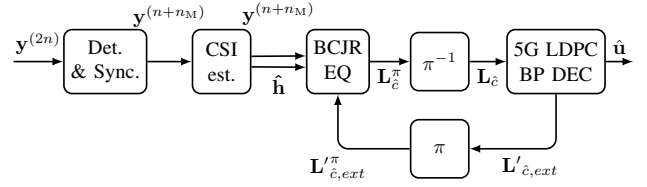
\begin{figure}[t]
	\centering
	\resizebox{0.48\textwidth}{!}{\input{tikz/baseline_RX_system_tikz}}
	\caption{Receiver model of a state-of-the-art conventional baseline system including preamble correlation, energy detection and channel estimation followed by iterative equalization, demapping and decoding between an BCJR equalizer and an BP decoder.}

	\label{fig:baseline_RX_system_tikz}
	\vspace*{-0.4cm}
\end{figure}

To address the task of transmitting $k$ bits within $n$ complex-valued channel uses, one can come up with a variety of different conventional systems.
We chose a single carrier short message system consisting of a Zadoff-Chu-based preamble sequence \cite{3gpp5G_phychannels} followed by an \ac{QPSK} modulated payload as a conventional baseline system for performance comparison.
For the specific parameters of $k=64$ bits within $n=64$ complex channel uses we empirically ended up with a preamble of length $n_\text{bl,ZF}=20$ symbols and a payload of $n_\text{bl,payload}=44$ symbols as best performing system.
An exemplary baseline system message is shown in Fig.~\ref{fig:baseline_blurb}.
Thus, the \ac{QPSK} modulated payload sequence conveys $k_\text{coded}=88$ coded bits.
To protect the $k=64$ information bits that have to be transmitted, we use a 5G NR compliant \ac{LDPC} code\footnote{Altough, Polar codes may yield a slightly better error-rate performance in the short length regime, the possibility of soft-output decoding and thereby a simpler integration in an \ac{IEDD} loop makes \ac{LDPC} codes a natural choice for the baseline.} \cite{5GLDPC} of rate $R_c = \nicefrac{64}{88} = \nicefrac{8}{11}$. The code is based on the second base graph with a lifting factor of $11$, where the first $22$ bits and the last $9$ bits are punctured and the bits $65$ and $66$ are shortened.

The receiver of the conventional baseline system is shown in Fig.~\ref{fig:baseline_RX_system_tikz} and consists of two separated steps, namely detection and decoding.
In a first step the detector tries to find a valid message within the received symbol vector $\yv^{(2n)}$ using a combination of preamble sequence correlation and energy detection.
If a certain threshold\footnote{This threshold has been empirically optimized for a specifically targeted false alarm rate of $0.1\%$ according to the 5GNR PRACH specifications \cite{3gpp5G_Basestation}} is reached, a message has been detected and the decoder forwards the estimated channel impulse response (which is the result of the preamble sequence auto-correlation) and the snippet $\yv^{(n+n_\text{M})}$ of the detected message to the BCJR equalizer.
This concludes the first step of message detection/synchronization and starts the second step of \acf{IEDD} \cite{Douillard1995IterativeCO}.
Therefore, a max-log BCJR equalizer approximates the \ac{MAP} \cite{erfanian94maxlog,bahl1974optimal,Robertson95suboptimalMAP} sequence estimate of the $k_\text{coded} = 88$  coded bits and forwards their \acp{LLR} $\mathbf{L}_{\hat{x}}$ to a  damped min-sum \ac{BP} decoder with a damping factor of $\lambda=0.7$.
The \ac{BP} decoder then decodes the codeword for $\ell_\text{BP}$ iterations and feeds the gained extrinsic information $\mathbf{L}_{\hat{x},\text{ext}}$ back to the BCJR equalizer.
While the BCJR equalizer could not make use of any \emph{a priori} information in the first iteration (as there has not been any feedback from the decoder yet, $\mathbf{L}_{\hat{x},\text{ext}} = \mathbf{0}$), it can now estimate the most likely bit sequence in form of $\mathbf{L}_{\hat{x}}$ again using the extrinsic information $\mathbf{L}_{\hat{x},\text{ext}}$ provided by the decoder.
The decoder feedback is considered in the equalizer by a weighted addition to the branch metrics.
However, the optimal weighting factor is a function of the \ac{SNR} which is assumed to be unknown.
Therefore, we empirically optimize the weight and set it to $0.2$. 
Note, we found that an \ac{SNR} estimation based on these short messages does not yield a gain in performance.
This process can be repeated for several $\ell_\text{IEDD}$ iterations resulting in a very competitive best-in-class conventional receiver system that can be considered state-of-the-art when it comes to bit-information retrieval at receiver side.
Finally, after $\ell_\text{IEDD}$ iterations, the decoder outputs the decoded bit sequence $\hat{\xv}$.

\section{Serial Turbo-Autoencoder-based System}
\label{sec:ae}

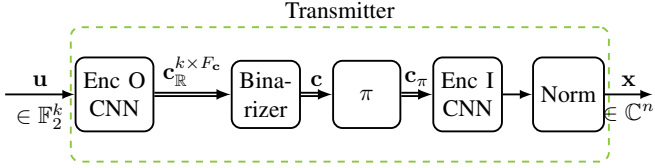
\begin{figure}[t]
	\centering
	\resizebox{0.5\textwidth}{!}{\input{tikz/serial_encoder}}
	\caption{The serially concatenated CNN-based encoder structure that embodies the transmitter.}

	\label{fig:serial_encoder}
	 \vspace*{-0.4cm}
\end{figure}

\begin{figure}[t]
	\centering
	\resizebox{0.5\textwidth}{!}{\input{tikz/serial_decoder}}
	\caption{CNN for message detection, followed by serial CNN-based Turbo-decoder structure for data retrieval.
	}

	\label{fig:serial_decoder}
	 \vspace*{-0.4cm}
\end{figure}
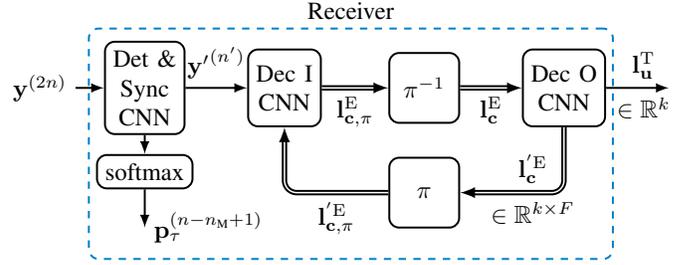

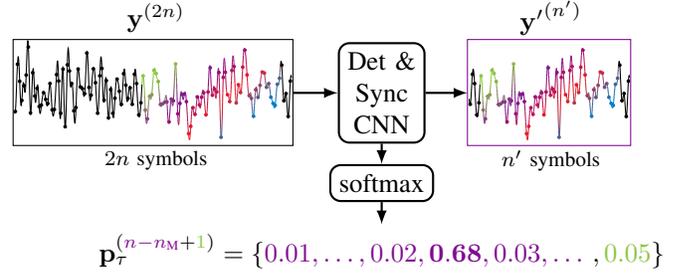
\begin{figure}[t]
	\centering
	\input{tikz/phyae_detector.tikz}
	\caption{Illustration of detection and synchronization with input $\yv^{(2n)}$ and outputs $\yv'^{(n')}$ and $\pv_\tau^{(n-n_\text{M}+1)}$.
	}
	\label{fig:phyae_detector}
	 \vspace*{-0.4cm}
\end{figure}

In this section, we introduce the proposed serial Turbo-autoencoder-based holistic short message system architecture for \ac{PHY}-level communication.
For the sake of convenience we will refer to this system as \emph{PHY-AE} in the following.
Fig.~\ref{fig:serial_encoder} shows the transmitter part of the PHY-AE which comprises piloting for detection and equalization as well as powerful \ac{FEC} encoding of payload information and modulation.
In Fig.~\ref{fig:serial_decoder} the two staged receiver part that comprises signal detection, synchronization, channel equalization and iterative decoding is shown.
Further, a detailed documentation of the employed \ac{CNN} architectures and hyperparameters is given in Appendix~\ref{sec:appendix}. %

\subsection{Transmitter Part}
The transmitter part, depicted in Fig.~\ref{fig:serial_encoder}, consists of the serial Turbo-autoencoder's encoder part and closely follows the architecture in \cite{clausius21serialAE} that showed, for the \ac{AWGN} channel, to be able to construct state-of-the-art short block codes.
Compared to this structure, the key adjustment to the encoder is a change of the final output activation to obtain a sequence of complex-valued channel uses $\xv^{(n)}$ as the encoder's output.
The main part stays the same, as the first encoder (outer encoder, \emph{Enc~O}) encodes the payload bits $\uv^{(k)}$ into a coded sequence $\cv_{\RR}$, where the subscript indicates that every entry is real-valued.
Then $\cv_{\RR}$ is binarized (using a saturated \ac{STE} \cite{BengioLC13, courbariaux2016binarized}) to $\cv \in \{-1,+1\}$ and $\cv$ is subsequently interleaved by a pseudo-random interleaver to $\cv_\pi$.
Finally, the inner encoder maps the coded sequence $\cv_\pi$ to the complex-valued symbol sequence $\xv^{(n)}$ which can then be transmitted over the channel.
Besides the power normalization, we constrain the message to zero mean.
We want to empathize that the autoencoder learns the holistic message structure solely based on the end-to-end applied loss without any guidance or injection of expert knowledge.

\subsection{Receiver Part}
\label{sec:phyae_rx}

The receiver structure is depicted in Fig.~\ref{fig:serial_decoder} and is also closely related to the receiver part of the serial Turbo-autoencoder \cite{clausius21serialAE}. %
However, the proposed structure includes the extension of a preceding \ac{CNN} to the Turbo-autoencoder's decoder to cope with the challenges of the random access scenario.
This preceding \ac{CNN} for detection and synchronization estimates the starting position of a message from a received sequence $\vec{y}^{(2n)}$, where the superscript indicates the length of the sequence.
Note that the detection can also predict that no message is contained in the received sequence.
If no message was detected, the subsequent \acp{CNN} for equalization and decoding can be omitted.
Otherwise, a snippet $\vec{y}^{(n')}$ with $n'=n+2n_\text{M}$ starting from $n_\text{M}$ positions before the predicted starting position is forwarded.
This process is schematically depicted in Fig.~\ref{fig:phyae_detector}.
The following \acp{CNN} are connected in an iterative fashion and separated by a deinterleaver and an interleaver.
The intuition is that the inner decoder learns a form of equalization/symbol detection with inherent \ac{CSI} estimation.
Since the a priori information from the outer decoder is subtracted from the output of the inner decoder, $\lv_{\cv}^{\mathrm{E}}$ can be interpreted as extrinsic information about the coded sequence $\cv$.
Subsequently, the outer decoder can calculate $\lv_{\uv}^\mathrm{T}$ which is a prediction of the transmitted sequence $\uv$.
Additionally, the outer decoder sends extrinsic feedback $\lv_{\cv}^{'\mathrm{E}}$ to the inner decoder about the code word sequence $\cv$.

Furthermore, a non-trivial extension to the serial Turbo-autoencoder structure is necessary in terms of the dimensions of the inputs and outputs.
In general, the dimensions of the exchanged information between inner and outer decoder can be chosen arbitrarily.
However, as we use one-dimensional \acp{CNN}, a suitable number of dimensions is 2.
We interpret the first dimension as a \emph{positional} dimension and the second dimensions as \emph{depth} with $F$ entries per position.
A practical simplification is to set the length of the positional dimension to $k$.
As a result, we can choose $F$ freely and concatenate inputs to a \ac{CNN} via the depth.
However, the a priori information and the synchronized channel observations do no share the length, but are of length $k=n$ and $n'$ respectively.
One way, that works well, is to align the lengths of the positional dimension via zero padding.
Note, that we set the number of bits $k$ and the number of channel uses $n$ to the same value ($k=n$).

\subsection{Training}

\begin{table}
	\caption{Hyperparameters for the training algorithm.}
	\label{tab:hyperparameter}
	\centering
	\begin{tabular}{l|l}
		Parameter& Value \\
		\hline
		Loss & BCE + CCE \\
		$T_{\mathrm{TX}}$  & $100$  \\
		$T_{\mathrm{RX}}$ &  $500$ \\
		Batchsize & $500-4000$  \\
		Optimizer & ADAM
	\end{tabular}
	\begin{tabular}{l|l}
		Parameter& Value \\
		\hline
		Learning rate & $10^{-4}-10^{-6}$  \\
		Encoder SNR & $15.0\mathrm{dB}$  \\
		Decoder SNR & $10.0-15.0\mathrm{dB}$ \\
		Detector SNR & $5.0-10.0\mathrm{dB}$ \\
		Loss-weight $\alpha$ & $0.01$
	\end{tabular}
\end{table}

We train the PHY-AE by optimizing the \ac{BCE} loss $L_\text{BCE,AE}$ between the originally transmitted input bit sequence $\uv$ and the estimated bit sequence $\lv_{\uv}^\mathrm{T}$ at the receiver's output.
In parallel, the detection CNN preceding the Turbo-decoder outputs a softmax activated prediction $\pv_\tau^{(n-n_\text{M}+1)}$ on the starting point of a detected message within the received sequence $\yv^{(2n)}$, as schematically shown in Fig.~\ref{fig:phyae_detector}.
On the basis of this prediction $\pv_\tau^{(n-n_\text{M}+1)}$ and the actual offset $\tau$ used by the channel model, we calculate a secondary \ac{CCE} loss $L_\text{CCE,det.}$ which is added to the \ac{BCE} loss.
The detection loss $L_\text{CCE,det.}$ can further be weighted by a hyperparameter $\alpha$ to balance decodability and detectability. Thus, the total loss is given by
\begin{equation}
    L_\text{tot.} = L_\text{BCE,AE} + \alpha \cdot L_\text{CCE,det.}.
\end{equation}

As a result, all components including encoder and decoder with preceding detection CNN contribute to the same loss $L_\text{tot}$ and are thereby trained in an end-to-end manner via \ac{SGD}.
This ensures the learning of holistic messages which are both optimized for detectability and decoding performance.
The training process itself is similar to \cite{jiang19Turboae,clausius21serialAE} and the main hyperparameters are given in Tab.~\ref{tab:hyperparameter}.
However, some adjustments have to be made to account for the random access scenario and the tapped-delay line channel model with memory.
First, the detector needs to be included in the alternating training process of updating the encoder without the decoder and vice versa.
To insure that the goals of the detector and the decoder are non-contradictory, the encoder weights are set to non-trainable during detector and decoder updates.
Yet, while keeping the alternating training schedule w.r.t. to the encoder/decoder training, the encoder (and thereby the learned messages) is jointly optimized for detection and decoding as the loss is calculated depending on both detector and decoder \acp{CNN}.
Note, that the trade-off parameter $\alpha$ also balances the contradicting goals of decodability and detectability in the encoder.
Another notable difference to the \ac{AWGN} case is, that the autoencoder \emph{over-fits} to the interleaver that is used during training.
As a consequence, the interleaver cannot be exchanged after training and must be chosen accordingly beforehand.
This behavior might be due to the memory induced disturbances in the channel and the resulting position-aware sequence estimation and tracking of the decoder.

Further, we want to emphasize that the BCE loss optimizes the BER and not, as often desired, the BLER.
As a result, we observed a significant number of block errors with only a single erroneous bit.
Thus, we propose a to add a single parity bit and flip the least reliable estimate in case the parity bit is not matching.
To account for the parity bit we chose to transmit one more bit, i.e., the autoencoder transmits $k_\text{PHYAE,actual}~=~k+1$ bits over $n=k$ channel uses. Please note, that this does not affect the final information rate $R$.
Further implementational details and the training procedure of a single epoch Alg.~\ref{alg:trainings_epoch} are given in Appendix~\ref{sec:appendix}.

\subsection{Discussion of Learned Messages and Power Limitation}

\begin{figure}
	\centering
	\input{tikz/hist_unconstrained.tikz}
	\caption{Heatmap histogram of the learned modulation (i.e., the constellation symbols of several thousand random messages, $k=64$, $n=64$) when the transmitter is only constrained by an average message power normalization.
	The green circles indicate where the PHY-AE has, implicitly, learned to place pilot symbols.}
	\label{fig:unlim_papr_scatter}
\end{figure}
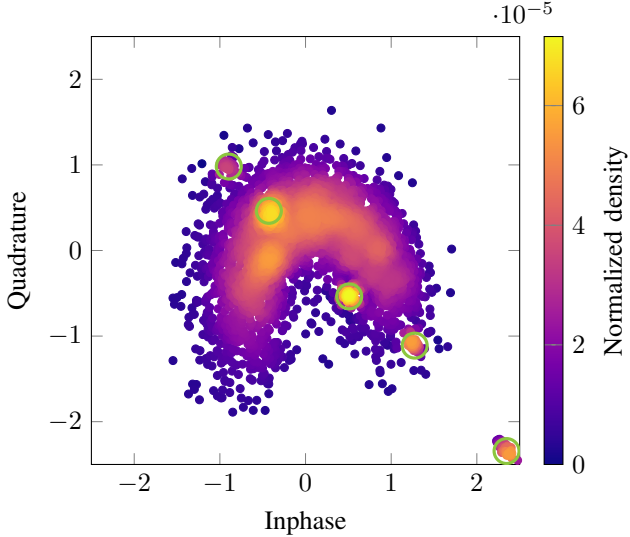

Before we take a look at the \ac{BER} and \ac{BLER} performance results, we first analyze the learned messages and discuss the impact of symbol power limitation with respect to the \acf{PAPR}.
Note, that it is not informative or even possible to inspect each learned message individually due to the sheer amount of $2^k$ differently modulated messages. Further, it is instructive to realize that the learned transmitter performs a non-linear encoding.
Thus, to get a better insight into the learned message sequences, Fig.~\ref{fig:unlim_papr_scatter} provides a scatter plot (heatmap histogram) of several thousand randomly generated messages, exemplary with parameters $k=64$ and $n=64$.
As can be seen, the transmitter learns a ``croissant''-shaped Gaussian distribution with one distinct position outside of the center at around $2.3-2.3j$, which is presumably used for piloting.
The fact that the learned distribution is not perfectly Gaussian or perfectly symmetrical indicates that a trade-off between detection/synchronization, channel estimation/equalization and information transfer was learned.
Within the learned distribution, we observe four more frequent positions (highlighted with green circles) which form an almost straight line together with the piloting position at $2.3-2.3j$.
We assume that symbols at these positions mainly serve the purpose of detection, synchronization and channel estimation and can thereby be seen as superimposed pilot positions, while the remaining positions within the ``croissant''-shape are most likely information carrying.
Interestingly, the observed shape of the distribution together with one distinct pilot position was reproducible in our experiments.
Starting with a random initialization, we trained the system multiple times from scratch and each time the learned messages exhibited a similarly shaped symbol distribution. %

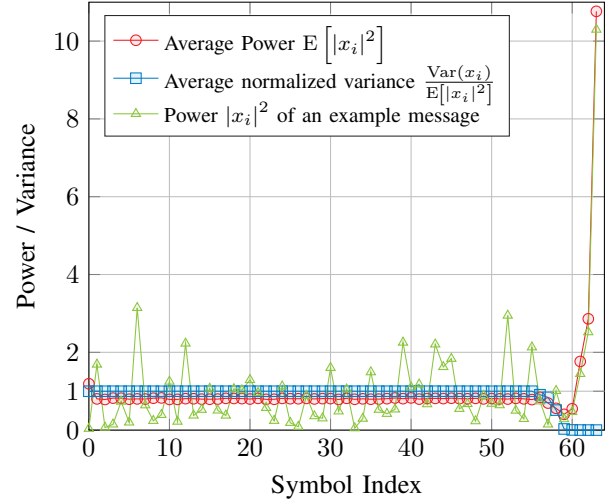
\begin{figure}[t]
	\centering
	\input{tikz/learned_const_energy.tikz}
	\caption{Power and variance distribution per symbol position $i$ of learned message sequences ($k=64$, $n=64$) when the transmitter is only constrained by an average message power normalization (no PAPR reduction applied here).}
	\label{fig:power_learned}
\end{figure}

Also if we take a look at the average power $\mathrm{E}\left[\left|x_i\right|^2\right] $ and variance $\mathrm{Var}\left(x_i\right)$ allocated to each symbol position $i$, as shown in Fig.~\ref{fig:power_learned}, we can clearly identify 5 distinct pilot symbols at the end of the learned messages.
These final 5 symbols do not carry significant information as their average variance is close to zero, meaning they consist of constant values just like traditional explicit pilot symbols.
Their average power also clearly differs from information carrying symbols.
Within the subset of investigated messages, the last symbol $x_{n-1}$ always corresponded to position $2.3-2.3j$, which is visible in Fig.~\ref{fig:unlim_papr_scatter}, and also the remaining 4 designated pilot symbols correspond to the more frequently used positions identified in Fig.~\ref{fig:unlim_papr_scatter} (green circles).
This result of the PHY-AE apparently learning to use a dedicated pilot sequence is of particular interest as it contradicts the information theoretical argument, that a dedicated preamble and payload are sub-optimal for detection and decoding \cite{lancho21jointdetdec,8461650}.
However, the task of the chosen scenario goes beyond slotted detection and decoding and also includes synchronization and channel estimation, which likely benefits from dedicated piloting symbols.
It is also not possible to determine if the PHY-AE solely relies on these dedicated pilots or if a superimposed piloting scheme is also present in the remaining symbols of a message.
Which can be assumed to some extent, due to the unsymmetrical non-Gaussian distribution of the remaining symbols.

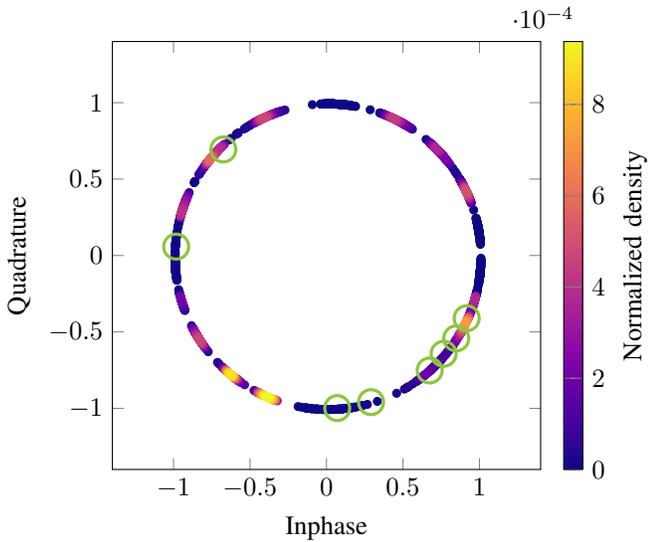
\begin{figure}
	\centering
	\input{tikz/hist_constrained.tikz}
	\caption{Heatmap histogram of the learned modulation ($k=64$, $n=64$) when the transmitter is constrained to a certain power range for each symbol (here $0.99 \leq |x_i|^2 \leq 1$).
	The green circles indicate where the PHY-AE has, implicitly, learned to place pilot symbols (we know these positions due to low variance $\mathrm{Var}\left(x_i\right)$).
	}
	\label{fig:const_mod_scatter}
\end{figure}

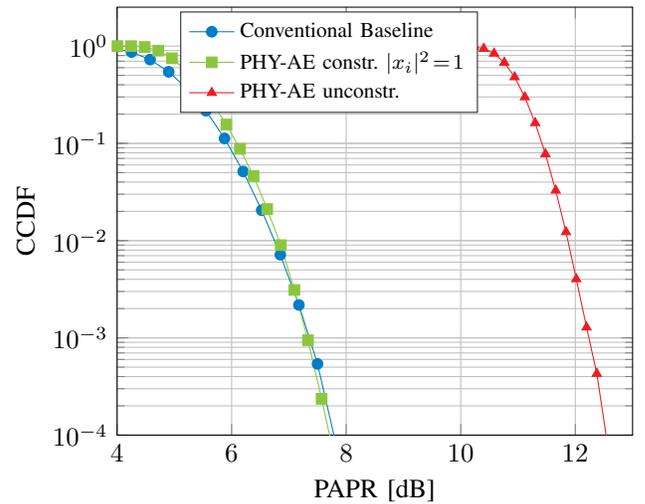
\begin{figure}[t]
	\centering
	\input{tikz/papr_ccdf.tikz}
	\caption{The \ac{CCDF} of the \ac{PAPR} in dB for the symbol power constrained and unconstrained PHY-AE.}
	\label{fig:cdf_h_len}
\end{figure}

Furthermore, we can see in Fig.~\ref{fig:power_learned}, that the transmitter learns to allocate a significant part of the available power per message to the last symbol $x_{n-1}$, presumably to easily measure the channel's impulse response.
Unfortunately, such high symbol power peaks lead to a large \ac{PAPR}, which is disadvantageous or even prohibitive for some hardware components, e.g., more expensive power amplifiers are required.
Especially with regard to cost- and power-efficient \acp{UE} either a limited \ac{PAPR} %
or even a constant envelope modulation can be required \cite{bockelmann16mmtc}.
For example the learned messages shown in Fig.~\ref{fig:unlim_papr_scatter}, which have only been constrained by an average power normalization (per message), exhibit a higher \ac{PAPR} than the conventional baseline system, as can be seen by the \ac{CCDF} of the PAPR depicted in Fig.~\ref{fig:cdf_h_len}.\footnote{As the \ac{PAPR} is defined for the actual analog signal that leaves the digital-to-analog converter after pulse shaping, we calculate a signal's \ac{PAPR} assuming \emph{sinc} pulse shaping (where \emph{sinc} pulse shaping is a worst case assumption in terms of \ac{PAPR}).}
However, the \ac{PAPR} can easily be reduced by limiting the individual symbol power $|x_i|^2$ to a desired range during training.
In the most extreme case the power of each symbol can thereby be limited to the unit circle, resulting in the symbol distribution depicted by Fig.~\ref{fig:const_mod_scatter}.
Consequently, the learned symbol power constrained PHY-AE messages shown in Fig.~\ref{fig:const_mod_scatter} entail an \ac{PAPR} reduction of more than 5 dB, as shown in Fig.~\ref{fig:cdf_h_len}, resulting in a PAPR performance comparable to that of the \ac{QPSK}-based conventional baseline.
As can be seen in the heatmap of Fig.~\ref{fig:const_mod_scatter}, the symbols of these messages are also not uniformly or symmetrically distributed over the unit circle, however, the assumed pilot positions (green circles), which we determined by their close to zero variance $\mathrm{Var}\left(x_i\right)$, are not the most frequently used positions anymore.
Yet, due to the reduction of the available modulation space to the unit circle, the density increases and positions are therefore used more often.
The consequences of symbol power limitation are, however, a more restricted modulation space that leads to less shaping gain, hindered channel estimation/equalization, and hindered detection, as will be shown in the next section.

\section{Results}
\label{sec:results}

We first evaluate the detection and decoding performance of the PHY-AE over the simulated channel model and, then, over-the-air within a real world \ac{SDR}-based deployment.
We compare the performance with the hand-crafted conventional baseline system introduced in Sec.~\ref{sec:baseline}.

\subsection{Detection Performance}

We define a detection error as the event where the detector is not able to detect and localize a message within a window of $n_\text{det}$ symbols to an accuracy of $\pm n_\text{M}$ symbols (misdetections \texttt{+} synchronization errors).
A detection error also occurs in the event of wrongly detecting a message where a ``none-message'' $\yv_\text{none}$, a recording of $n_\text{det}$ symbols without any transmitted message, was simulated (false alarm).
As a decision for a threshold cannot optimize both types of errors jointly, we set the threshold such that the false alarm rate is lower than $0.1\%$ according to the 5G NR specifications for the \ac{PRACH} \cite{3gpp5G_Basestation}.
This has been ensured empirically by simulations to find the optimal threshold for the conventional baseline system and by adjusting $\alpha$ and the amount of none-messages during training of the PHY-AE.

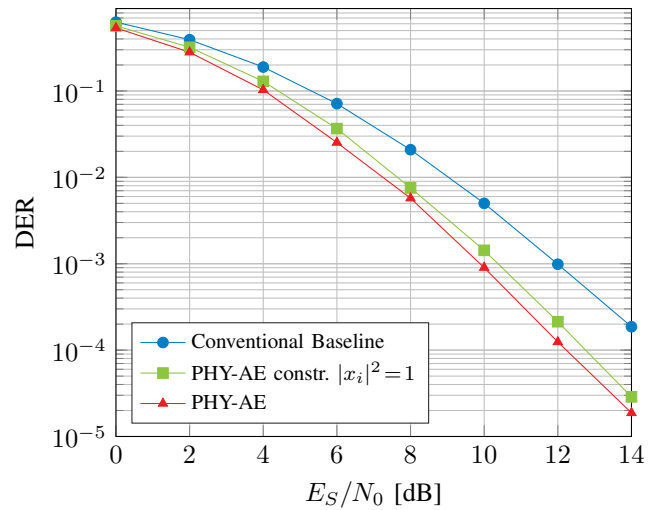
\begin{figure}[t]
\centering
\input{tikz/der_prach.tikz}
\caption{Simulated \acf{DER} (detection errors: misdetections \texttt{+} synchronization errors) vs. \ac{SNR} for the channel model with $n_\text{taps}=5$ random channel taps, $k=64$, $n=64$ and $n_\text{det}=128$.}
\label{fig:der_sim}
\end{figure}

As a result, Fig.~\ref{fig:der_sim} shows the simulated \ac{DER} for the conventional baseline system and the PHY-AE systems.
As can be seen, the unconstrained \emph{PHY-AE} shows the best \ac{DER} performance.
We assume that the possibility to allocate more energy into single symbols enables better detection.
Yet, also the maximal symbol power-constrained \emph{PHY-AE constr.} with $|x_i|^2\!=\!1$ significantly outperforms the correlation and energy detection-based conventional baseline.
The overall better detection performance of the PHY-AE's CNN-based detector shows the advantage of a system that is optimized for both data transmission \emph{and} detection from end-to-end. %

\subsection{Decoding Performance}

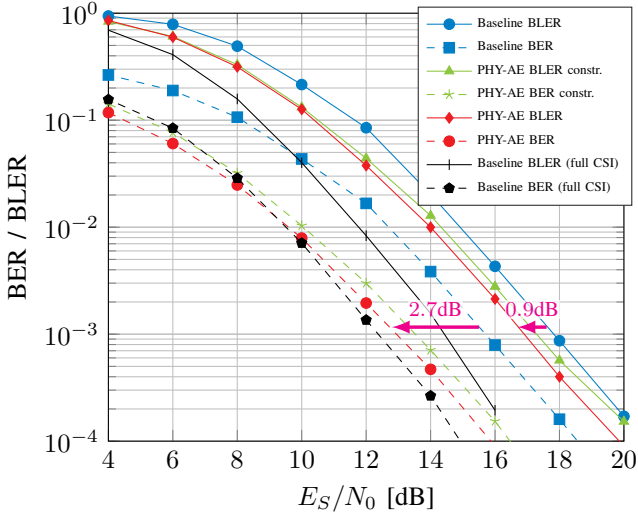
\begin{figure}[t]
	\centering
	\input{tikz/ber_prach_toff_sim.tikz}
	\caption{Simulated BLER (solid) and BER (dashed) vs. SNR over the channel model with $n_\text{taps}=5$ random channel taps ($k=64$, $n=64$)}
	\label{fig:ber_sim_toff}
\end{figure}

After successful detection of a received message, the receivers needs to equalize $\yv'$ and estimate the originally transmitted bit sequence $\uv$.
Fig.~\ref{fig:ber_sim_toff} shows the achieved BLER and BER performance of the conventional \emph{Baseline} system, the unconstrained \emph{PHY-AE} and the maximal symbol power constrained ($|x_i|^2\!=\!1$) \emph{PHY-AE constr.} for parameters $k=64$ and $n=64$.
As a reference, we also show the maximal achievable performance \emph{Baseline (full CSI)} of the baseline system where the \ac{CSI}, i.e., $\hv$ and $\tau_{\mathrm{STO}}$, is perfectly known at the receiver side.
One can observe that the previously discussed DER of all systems (Fig.~\ref{fig:der_sim}) is $\approx \! 6$dB better than their respective BLER.
This means that detection errors have almost no impact on the overall performance for the simulated channel.
We can also see that the \emph{PHY-AE} performs up to 1dB better in terms of BLER and more than 2dB better in terms of BER compared to the conventional \emph{Baseline}.
One particularly interesting observation is that the symbol power constrained \emph{PHY-AE constr.} is almost on par with the unconstrained \emph{PHY-AE}, which shows that symbol power limitation only results in slightly degraded end-to-end performance.
Another interesting result is that the \emph{PHY-AE} even outperforms the \emph{Baseline (full CSI)} with perfect \ac{CSI} in terms of BER in the low SNR regime.
These large BER gains indicate that the \emph{PHY-AE} must have learned a holistic coding scheme with notable coding gain considering the short block length of $k=64$.

\subsection{Universality and Scope of Application}

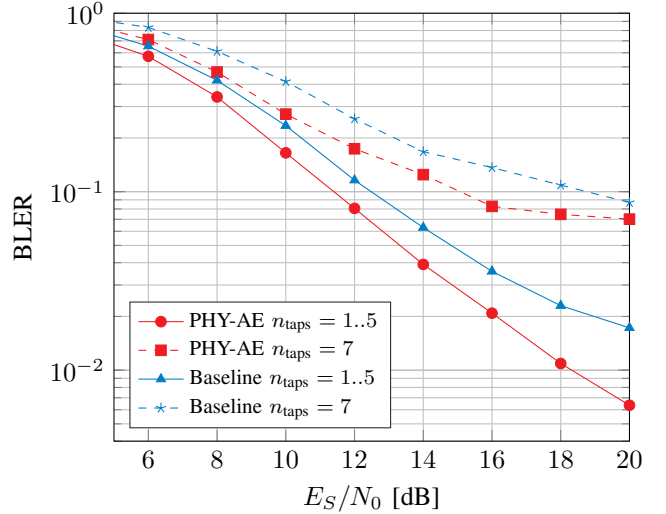
\begin{figure}[t]
	\centering
	\input{tikz/ber_prach_sim_h.tikz}
	\caption{Simulated BLER vs. SNR over the channel model with a varying number of $n_\text{taps}=1..5$ (solid) and $n_\text{taps}=7$ (dashed) random channel taps ($k=64$, $n=64$), while both systems are optimized for $n_\text{taps}=5$.}
	\label{fig:ber_sim_h}
\end{figure}

A common misconception in terms of model-based training is that NN-based PHY-layer components optimized over a channel model will only work for the exact same channel parameters.
However, if the channel model relies on stochastically drawn parameters, it covers an ensemble of many different scenarios.
In the case of the chosen channel model described in Sec.~\ref{sec:channel_model}, where the channel taps $\hv$ are randomly drawn, the PHY-AE also encounters simpler channel instantiations with effectively $n_\text{taps} \leq 5$ channel taps during training (as some instantiations of $h_i$ can be close to zero).
From this perspective, the employed channel model acts more like a hyperparameter of the PHY-AE itself, as it defines an upper design limit for the PHY-AE, which learns a robust signaling for up to $n_\text{taps}=5$ channel taps, but can also operate under simpler conditions.
This is similar to the process of designing the length of a dedicated preamble for a conventional system with respect to the desired scope of application.
Fig.~\ref{fig:ber_sim_h} therefore depicts the averaged \ac{BLER} performance of the PHY-AE and the baseline system over the channel model with explicitly reduced numbers of simulated channel taps from $n_\text{taps}=1,\dots,5$.
As can be seen, the PHY-AE also outperforms the conventional baseline over simulated channels with $n_\text{taps} \leq 5$ channel taps.\footnote{The differing slope and shifted BLER performance visible in Fig.~\ref{fig:ber_sim_h} compared to the results for $n_\text{taps}=5$ shown in Fig.~\ref{fig:ber_sim_toff} are due to a higher SNR variance for simulated channel instantiations with $n_\text{taps}<5$.}
Fig.~\ref{fig:ber_sim_h} also depicts the \ac{BLER} performance of the PHY-AE and the baseline system for $n_\text{taps}=7$ channel taps (dashed lines), which is out of the scope of application for both systems.
As expected, both systems show a degraded BLER performance for $n_\text{taps}=7$ while the PHY-AE still outperforms the baseline system, indicating a generalization to the task.
To be able to operate over channels with up to $n_\text{taps}=7$ channel taps, the PHY-AE would require re-training and the baseline system would have to be redesigned w.r.t. preamble length and payload encoding.

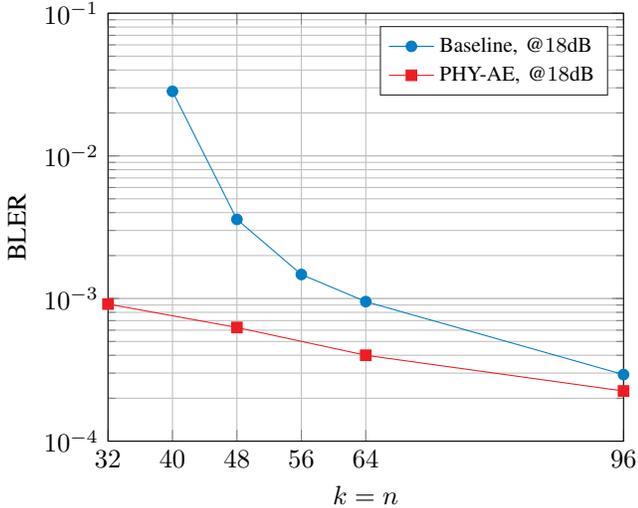
\begin{figure}[t]
	\centering
	\input{tikz/ber_length_sweep.tikz}
	\caption{Simulated \ac{BLER} vs. block length $n$ evaluated at an SNR of $18$dB and fixed information rate of $R = \nicefrac{k}{n} = 1$ for each block length $n$.}
	\label{fig:ber_length_sweep}
\end{figure}

To provide a better understanding of the general superiority of learned holistic messages and joint CNN-based detection, synchronization, equalization and decoding in the field of short-packet communications, Fig.~\ref{fig:ber_length_sweep} shows the BLER evaluated at a fixed SNR of $18$dB  for various block lengths $n$.
For this evaluation the PHY-AE has been retrained and the handcrafted baseline has been empirically redesigned for each $n$ to transmit $k=n$ information bits.
Further implementational details are given in Appendix~\ref{sec:appendix}.
As can be seen, the PHY-AE outperforms the conventional baseline for all simulated block lengths up to $n=96$.
Especially for shorter block lengths the gain of the PHY-AE compared to the baseline further increases.
On the one hand, this is linked to the practical problem of how to reasonably design a conventional system with explicit preamble and payload allocation for ultra short block lengths, as short as $n=32$.
On the other hand, this result can be easily explained by the information theoretical advantages of holistic messages and joint processing over provably sub-optimal dedicated preambles \cite{lancho21jointdetdec} for short block lengths.
Another observation is that the performance of both systems increases with the block length $n$ and the number of transmitted information bits $k$, which is also expected as it can be explained by the increasing achievable coding rate for increasing $k$ \cite{Polyanskiy2010}.
Yet, we know from \cite{clausius21serialAE}, that the coding gain of the Turbo-autoencoder architecture does not scale as well as for conventional coding schemes with increasing $k$ for $k \gg 100$.
Therefore, we expect for increasing $n$, as soon as detection and estimation can be handled by a sufficiently long preamble sequence and coding gain predominates, that a conventional system will, at some point, surpass the PHY-AE.
However, this is not the main focus of this work as for long block lengths of $n \gg 100$ the disadvantages of a dedicated preamble vanish and conventional systems with state-of-the-art coding schemes are already operating close to capacity.

\subsection{Actual Deployment and Measurement Testbed}
\label{sec:ota_deployment}

To demonstrate the applicability of the proposed system, we deploy the PHY-AE on an \ac{SDR}-based testbed and evaluate the \ac{OTA} performance on a real world channel.
Therefore, we fix the weights of the transmitter part, which had been optimized via end-to-end \ac{SGD} updates through the stochastic channel model.
Although it was shown in \cite{Aoudia2019} and \cite{doerner2020gan}, that it is also possible to further optimize the transmitter through the actual channel using \ac{RL} methods or, respectively, a \ac{GAN}-based channel model, we choose to not further adapt the transmitter signaling.
We see multiple motivations behind this decision.
First, re-optimizing the transmitter means a potential loss of generalization as it could lead to overfitting to a certain dataset or channel condition.
To counter (or leverage) overfitting, periodical \ac{RL}-based re-training steps would be required to \emph{follow} changing channel conditions throughout the whole topological area of operation and different environments.
A similar problem holds for the \ac{GAN}-based approach, where a dataset of channel conditions for the entire scope of application is required to be able to train a generative model without overfitting effects.
Secondly, training the transmitter weights in-the-field would require significant computational power at the transmitter devices, which contradicts a scenario of low-power \ac{MTC} devices.
And thirdly, both transmitter training approaches would require some kind of feedback channel to the transmitter devices to update their weights, which requires additional bandwidth and also contradicts an application scenario of massive amounts of ultra low-budget \emph{transmit-only} devices.
This is why we deploy the transmitter weights learned with the stochastic channel model described in Sec.~\ref{sec:channel_model}, which was deliberately designed to generate harsh channel conditions for the targeted scope of application.

At the receiver side, on the other hand, we also show the performance of a finetuned receiver according to \cite{Doerner2018}.
As has been shown in \cite{Doerner2018} and \cite{cammerer2019tcom}, the performance loss due to unavoidable mismatch between the synthetic channel model used for training and the actual channel can be significantly mitigated by finetuning the receiver.
We used $12.5 \cdot 10^6$ recorded messages and their corresponding genie labels to finetune the receiver via re-training.
However, in practice the required labels for the receiver re-training process can also be easily obtained by the periodical transmission of pilot messages or by leveraging the idea of \ac{ECC}-based label recovery formulated in \cite{schibisch2018online}.

\begin{figure}[t]
	\centering
	\input{tikz/channel_h.tikz}
	\caption{Averaged channel impulse response during \ac{OTA} measurements at 2.35GHz and 40.0MHz bandwidth with the \ac{SDR}-based testbed.}
	\label{fig:channel_h}
\end{figure}
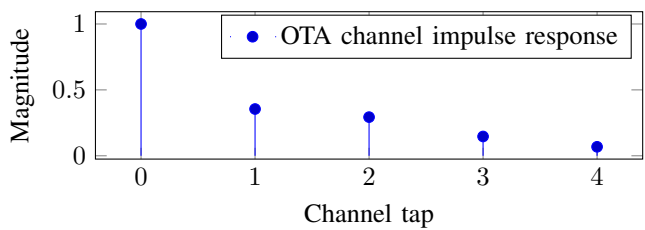

Our measurement testbed consists of two \acp{USRP} B210 from Ettus Research, transmitting at a carrier-frequency of 2.35GHz and a bandwidth of 40.0MHz.\footnote{As our testbed was optimized for carrier frequencies in the WiFi spectrum, we had to use a high bandwidth to experience notable delay spread. The goal was to mimic similar conditions that \ac{IoT} devices find at typically used unlicensed bands $\sim \! 900$MHz for bandwidths of $\sim \! 200$kHz.}
The \acp{USRP} are placed in a static indoor office environment without line of sight and the measured channel impulse response is depicted in Fig.~\ref{fig:channel_h}.
Transmitter and receiver USRP are not clock-synced and during measurements we rely on each USRP's internal oscillator clock with a declared frequency accuracy of $\pm2.0$ ppm. %
Except the USRP's internal hardware digital to analog converters we do not use over-sampling at the transmitter as well as at the receiver side, i.e., $\xv$ is directly fed to the transmitter USRP and $\yv$ is recorded at the same sampling frequency at the receiver USRP.

\subsection{Over-the-Air Results}
\label{sec:ota_results}

\begin{figure}[t]
\centering
\input{tikz/der_prach_real_ch.tikz}
\caption{\ac{DER} (misdetections \texttt{+} synchronization errors) vs. \ac{SNR} for over-the-air measurements with the SDR testbed, $k=64$, $n=64$ and $n_\text{det}=128$.}
\label{fig:der_ota}
\end{figure}
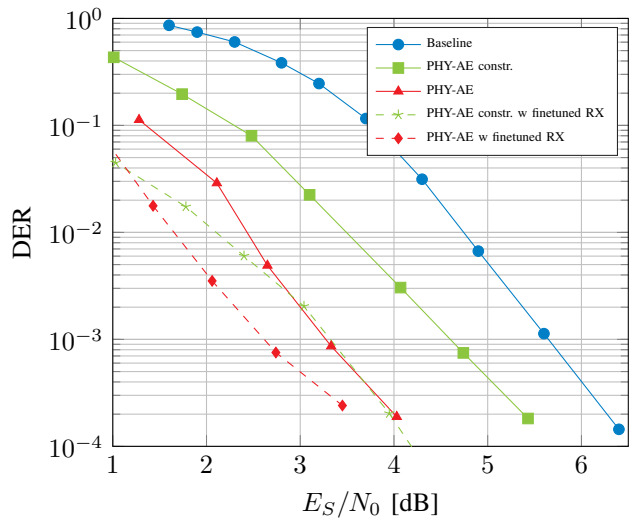

As with the simulated results before, we first have a look at the detection error rate for over-the-air measurements with the SDR testbed.
Fig.~\ref{fig:der_ota} shows the OTA DER for the conventional \emph{Baseline} system, the $|x_i|^2\!=\!1$ symbol power constrained \emph{PHY-AE constr.} and the unconstrained \emph{PHY-AE}.
It also shows the performance of the PHY-AE with a finetuned receiver.
As a first important observation we can see that both variants \emph{PHY-AE} and \emph{PHY-AE constr.}, which have only been trained and optimized for the synthetic channel model, perform -- out-of-the-box -- better than the correlation and energy detection-based conventional \emph{Baseline}.
Similar to the simulated results, we can also see that the unconstrained \emph{PHY-AE} achieves an even better performance than the symbol power constrained \emph{PHY-AE constr.} variant.
If we then finetune the receiver of the PHY-AE via re-training to further fit the NN to all measurement and hardware impairments of the SDR testbed, we can see a significant performance gain of about 0.5dB for the \emph{PHY-AE w finetuned RX} and about 1dB for the constrained \emph{PHY-AE constr. w finetuned RX}.
Based on these results we can state that the PHY-AE is able to confirm its superior detection performance over-the-air (despite the channel mismatch) and the PHY-AE with a retrained receiver can outperform the conventional system even more due to adaptation to hardware insufficiencies.

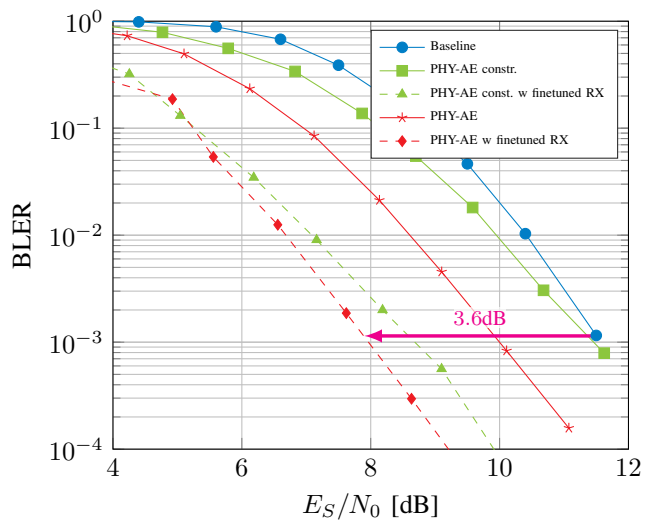
\begin{figure}[t]
\centering
\input{tikz/ber_prach_real.tikz}
\caption{\ac{BLER} vs. \ac{SNR} for over-the-air measurements with the SDR testbed, $k=64$, $n=64$. 
}
\label{fig:ber_ota}
\end{figure}

When looking at the OTA BLER performance shown in Fig.~\ref{fig:ber_ota}, we can see similar results.
Both the unconstrained \emph{PHY-AE} and the $|x_i|^2\!=\!1$ constrained \emph{PHY-AE constr.} show a better BLER performance compared to the \emph{Baseline} system out-of-the-box.
This shows, again, that the chosen synthetic channel model is accurate enough to enable the PHY-AE to generalize to real world channels.
If we further finetune the PHY-AE's receiver part, we can again observe a significant performance gain of about 2-2.5dB for both variants, while the constrained PHY-AE seems to profit slightly more from finetuning.
Note, that finetuning in this case also means deliberate overfitting to our static measurement environment, and for real world application periodic finetuning steps would be necessary to follow a changing environment \cite{fischerAdaption2022}.

In conclusion to these measurement results we can state that the PHY-AE is as reliable as the handcrafted state-of-the art conventional system for $n=k=64$, but already at up to 3.6dB lower SNR.

\section{Outlook and Conclusion}
\label{sec:conclusion}

We have proposed an end-to-end optimized autoencoder-NN-based solution for joint detection, synchronization, equalization and decoding in short-packet communications.
Our proposed system shows better DER, BER and BLER performance than a handcrafted state-of-the-art conventional baseline system over a simulated multipath channel model for block lengths of up to $k \! \sim \! 100$ information bits.
We further demonstrated the applicability of the PHY-AE in actual over-the-air measurements within an \ac{SDR}-based testbed, where the superior performance was confirmed and gains were further increased by additional receiver finetuning.
Due to improved performance in the domain of short-packet communications and the conceptual simplicity of learned holistic messages, we think this approach renders a viable option for future \ac{mMTC} systems and, potentially, also for initial channel access schemes.

Several open questions remain, e.g.:
a) It would be advantageous to extend the end-to-end loss function to explicitly optimize the (often) desired BLER metric instead of the BER via BCE.
b) Adaptive finetuning of transmitter and receiver for changing channel conditions could be an interesting investigation.
And c) also an extension of the PHY-AE to waveform-level signal modulation appears to be straightforward as it would allow for direct PAPR and \ac{ACLR} targets during training \cite{AoudiaWaveform2022}.

\bibliographystyle{IEEEtran}
\bibliography{IEEEabrv,references}

\newpage

\appendix{
\label{sec:appendix}

\begin{table}
	\caption{Hyperparameters of the \acp{CNN}}
	\label{tab:hyperparameter_cnn}
	\centering
	\begin{tabular}{l|l}
		Parameter& Value \\
		\hline
		Num. filters & $100$ \\
		Layers per \ac{CNN}  & $5$  \\
		Kernel &  $5$ \\
	\end{tabular}
	\begin{tabular}{l|l}
		Parameter& Value \\
		\hline
		Activation &  ELU \\
		Decoder Iterations & $6$  \\
		$F_{\cv} = F$ & $10$   \\
	\end{tabular}
\end{table}

\begin{figure}[t]
	\centering
	\resizebox{0.5\textwidth}{!}{\input{appendix/cnn_block.tikz}}

	\caption{Flowchart of the basic \ac{CNN} block.
	}

	\label{fig:basic_cnn}
	 \vspace*{-0.4cm}
\end{figure}
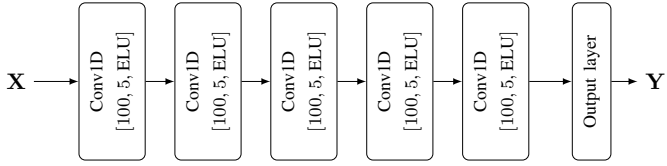

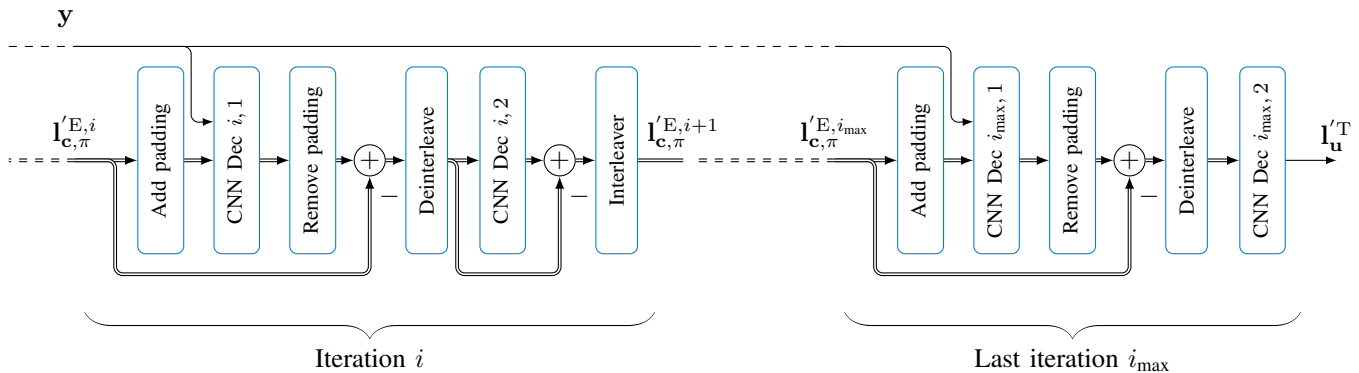
\begin{figure*}[t]
	\centering

	\resizebox{1.0\textwidth}{!}{\input{appendix/generator_structure.tikz}}
	
	\caption{Unrolled decoder with implementation details. In the first iteration the a priori information is set to  $\lv_{\cv,\pi}^{'\mathrm{E},1}=\mathbf{0}$. Double arrows indicate a channel depth of size $F$.
	}

	\label{fig:unrolled_decoder}
	 \vspace*{-0.4cm}
\end{figure*}

\begin{algorithm}[t]
\caption{Training procedure of one epoch for the \emph{PHY-AE}.
}
	\SetAlgoLined
	\SetKwInOut{Input}{Input}
	\SetKwInOut{Output}{Output}
	\SetKwBlock{Repeat}{repeat}{}
	\SetKwFor{RepTimes}{For}{do}{end}
	\DontPrintSemicolon
	\Input{$\gamma, \alpha, T_{\mathrm{TX}}, T_{\mathrm{RX}}, \theta_\text{ENC},  \theta_\text{DET},\theta_\text{DEC}, \sigma^2_\text{ENC}, $\\$ \sigma^2_\text{DET,min},\sigma^2_\text{DET,max}, \sigma^2_\text{DEC,min},\sigma^2_\text{DEC,max}$}
	\Output{$\theta_\text{ENC},  \theta_\text{DET},\theta_\text{DEC}$}
	\BlankLine
	\tcp{Transmitter training}
	\RepTimes{$i=1,\dots,T_{\mathrm{TX}}$}{
	    set\_trainable$(\theta_\text{ENC},\theta_\text{DET},\theta_\text{DEC}) =[\text{True, False, False}]$\\
	
	    $\uv \gets \text{generate\_Bits}()$ \\

	   $\xv \gets \text{encode}\left(\uv ;\theta_\text{ENC} \right)$ \\
	   
	   $\yv \gets \text{transmit}\left(\xv, [\sigma^2_\text{ENC}] \right)$ \\

	   \tcp{Detector with $\sigma^2_\text{ENC}$}
	   $\mathbf{p}_{\mathbf{\tau}} \gets \text{detection}\left(\yv\right)$ \\

	    $L_{\text{CCE,det}} \gets \text{CCE}\left(\tau_{\text{off}}, \mathbf{p}_{\mathbf{\tau}} ;\theta_\text{ENC} \right)$\\

	   \tcp{Decoder with $\sigma^2_\text{ENC}$}
	   $\hat{\tau}_{\text{off}}\gets \text{argmax}\left(\mathbf{p}_{\mathbf{\tau}}\right)$ \\
	   $\yv_{\text{cutout}}\gets \text{cutout}\left(\yv ,\tau_{\text{off}} \right)$ \\
	   
	    $\lv_{\uv}^\mathrm{T}\gets \text{decode}\left(\yv_{\text{cutout}}\right)$ \\
	    $L_{\text{BCE,dec}} \gets \text{BCE}\left(\uv, \lv_{\uv}^\mathrm{T} \right)$\\

	   $L_{\text{tot}} \gets  L_{\text{BCE,dec}}+ \alpha \cdot L_{\text{CCE,det}} $ \\
		$\theta_\text{ENC} \gets \text{SGD}\left(\theta_\text{ENC}, L_{\text{tot}}\right)$
	}
	\tcp{Receiver training}
	\RepTimes{$i=1,\dots,T_{\mathrm{RX}}$}{
    	set\_trainable$(\theta_\text{ENC},\theta_\text{DET},\theta_\text{DEC}) =[\text{False, True, True}]$\\
    	$\sigma^2_\text{DET},\sigma^2_\text{DEC} \gets \text{generate\_SNR}([[\sigma^2_\text{DET,min},\sigma^2_\text{DET,max}] ,$\\~$[\sigma^2_\text{DEC,min},\sigma^2_\text{DEC,max}]])$

	   $\xv \gets \text{encode}\left(\uv ;\theta_\text{ENC} \right)$ \\
	   
	   $\yv_{\text{DET}}, \yv_{\text{DEC}} \gets \text{transmit}\left(\xv, [\sigma^2_\text{DET}, \sigma^2_\text{DEC} ] \right)$ \\
	   
	    \tcp{Detector with SNR $\sigma^2_\text{DET}$}
	    $\text{None-messages} \gets \text{generate\_none\_msg}(\gamma)$\\
	    $\yv_{\text{DET}}\gets \text{concat} \left( \yv_{\text{DET}},\text{None-messages} \right)$ \\
	    $\mathbf{p}_{\mathbf{\tau}} \gets \text{detection}\left(\yv_{\text{DET}} ;\theta_\text{DET} \right)$ \\
	    $L_{\text{CCE,det}} \gets \text{CCE}\left(\tau_{\text{off}}, \mathbf{p}_{\mathbf{\tau}} ;\theta_\text{DET}\right)$\\
	    \tcp{Decoder with SNR $\sigma^2_\text{DEC}$}
	    $\hat{\tau}_{\text{off}}\gets \text{argmax(detection}\left(\yv_{\text{DEC}}  \right) ) $ \\
	    $\yv_{\text{cutout}}\gets \text{cutout}\left(\yv_{\text{DEC}} ,\hat{\tau}_{\text{off}} \right)$ \\
	    $\lv_{\uv}^\mathrm{T}\gets \text{decode}\left(\yv_{\text{cutout}} ;\theta_\text{DEC} \right)$ \\
	    $L_{\text{BCE,dec}} \gets \text{BCE}\left(\uv, \lv_{\uv}^\mathrm{T}  ;\theta_\text{DEC}\right)$\\
	    $L_{\text{tot}} \gets  L_{\text{BCE,dec}}+ \alpha \cdot L_{\text{CCE,det}} $ \\
        $\theta_\text{DET}, \theta_\text{DEC} \gets \text{SGD}\left([\theta_\text{DET}, \theta_\text{DEC}], L_{\text{tot}}\right)$
	}
	\tcp{Adapting \emph{None}-message ratio}
    \uIf{$\text{\acs{FAR}} \geq 0.1\%$}{
                $\gamma \gets \gamma +0.05$\\
			}
	\uElse{$\gamma \gets \gamma -0.05$\\}
	
	\label{alg:trainings_epoch}
\end{algorithm}

In the appendix we want to provide further implementation details of the autoencoder.
All following parts of the system are based on a basic \ac{CNN}-structure as shown in Fig.~\ref{fig:basic_cnn}.
This structure consists of $5$ one-dimensional convolution-layers with \emph{same}-padding that are parameterized by the triple $\left[ N_{\text{filters}},L_{\text{kernel}}, f_{\text{activation}} \right]$ which is the number of filters, the length of the kernel and the used activation function respectively.
The output layer is chosen according to the use case within the PHY-AE system.
Further, Alg.~\ref{alg:trainings_epoch} shows a more detailed (but probably still incomplete) summary of the training process. %

\subsection{PHY-AE Encoder}
The output layer of the outer encoder is a one-dimensional convolutional-layer with the hyperparamters $\left[ N_{\text{filters}},L_{\text{kernel}}, f_{\text{activation}} \right]$ = $\left[ 10,1, \text{None} \right]$.
The task of the layer is to combine the channels per position to the desired output of $F=10$ coded features per bit-position.
Similarly, the output layer of the inner encoder is a one-dimensional convolutional-layer with the hyperparamters $\left[ N_{\text{filters}},L_{\text{kernel}}, f_{\text{activation}} \right]$ = $\left[ 2,1, \text{None} \right]$. In this case the channels are combined to resemble the real part and imaginary part of a complex symbol.
The final processing step in the transmitter concerns the normalization. The basic normalization step ensure that each symbol sequence $\xv$ has a zero mean with unit energy via
\begin{equation}
    \xv = \frac{\xv'-\mu_{\xv'}}{\sigma_{\xv'}}
\end{equation}
where $\mu_{\xv'}$ is the mean and $\sigma_{\xv'}$ the standard deviation of $\xv'$.
Furthermore, during inference of the symbol power constrained PHY-AE, we only consider the phase of the symbol and set the magnitude to $|x_i|=1$.
While during training, we only clip $x_i$ according to a power-threshold per symbol to ensure a stable training process, instead of considering the phase.
To achieve convergence we successively decrease this power-threshold to $|x_i|^2=2$, $|x_i|^2=1.1$ and $|x_i|^2=1.01$. %
For each threshold we train until the test-loss saturates.

\subsection{Detector}
Before we describe the output layer of detector-\ac{CNN} in detail, we want to focus on the input to the network.
A necessary pre-processing step is some form of normalization to account for different received power levels.
For this, we chose to normalize the channel observations $\yv$ to unit power.
Otherwise a varying transmit power would be required during simulation.

The output layer of the detector actually contains three layers.
The first layer is a one-dimensional convolutional layer with hyperparameters $\left[ N_{\text{filters}},L_{\text{kernel}}, f_{\text{activation}} \right]$ = $\left[ 1,1, \text{None} \right]$ to combine the channels.
The second layer is a flatten layer and the third layer is a dense layer with $n-n_{\text{M}}+1$ neurons followed by a softmax activation function.
Due to the softmax activation, each entry of $\pv_\tau^{(n-n_\text{M}+1)}$ can be interpreted as a probability of a specific integer offset $\tau_{\text{off}} =  [0,1...,n-n_{\text{M}}]$ and the final entry is the estimate whether a message is present in the observed sequence or not. 

For the training of the detector we use a \ac{CE}-loss.
As the detector-\ac{CNN} needs to find a trade-off between \acp{MD} and \acp{FA} we can split the loss function in two terms.
Note, that errors due to wrong synchronization are interpreted as a \ac{MD}.
Thus, the loss can be rewritten as
\begin{equation}
    L_{\text{CE}}= L_{\text{CE,MD}} + \gamma \cdot L_{\text{CE,FA}}
\end{equation}
where $L_{\text{CE,MD}}$ is the loss due to a \ac{MD}, $\gamma$ is the ratio \emph{none}-messages in the dataset and $L_{\text{CE,FA}}$ is the loss due to a \ac{FA}.
Therefore, we can target a specific \ac{FAR} by choosing $\gamma$ accordingly. 
For this, we propose to adapt $\gamma$ during training based on the \ac{FAR}. 
At the end of each epoch we check whether the \ac{FAR} is smaller or larger than the target \ac{FAR} and increase or decrease $\gamma$, respectively.  

Further, we want to emphasize that we use different SNRs for detection and decoding during training, since the interesting SNR region for detection is roughly $6$dB lower than for decoding.

\subsection{Decoder}
While Fig.~\ref{fig:serial_decoder} already shows the iterative structure of the decoder, two implementation details are left out.
Both are shown in Fig.~\ref{fig:unrolled_decoder}.
The first one regards the padding of the a priori information.
In order to concatenate the a priori information with the channel observations we pad it from the left and the right with $2n_\text{M}$ zeros in total.
The second detail concerns the calculation of the extrinsic information. 
It is calculated by subtracting the a priori information from the output of a decoder-\ac{CNN}.
Conveniently, this acts as a residual connection in the network that might facilitate the training process. 
}

\subsection{Achieving Variable Message Lengths}

In order to achieve the BLER results for various message lengths $n$, as shown in Fig.~\ref{fig:ber_length_sweep}, we had to retrain the PHY-AE and slightly re-design the conventional baseline system.
The PHY-AE was retrained from scratch for all lengths $n$, except for $n=32$ as we examined that a retrained $n=64$-variant performed better at $n=32$ than the $n=32$-variant learned from scratch.
For the conventional baseline, we empirically optimized the trade-off between pilots and payload w.r.t. each message length $n$. 
Thus, we still use QPSK modulation and a 5G \ac{LDPC} code with the respective length and rate.
As a result, we chose $n_\text{bl,ZF}=16$ pilot symbols for $n=40$ and $n=48$, $n_\text{bl,ZF}=20$ for $n=56$ and $n=64$, and finally $n_\text{bl,ZF}=24$ for $n=96$.
Especially for short lengths $n<56$, reducing the number of pilot symbols led to further performance degradation of the baseline system as can be seen in Fig.~\ref{fig:ber_length_sweep}.

\end{document}

%% file: macros.tex
\renewcommand{\vec}[1]{\mathbf{#1}}

\newcommand{\cv}{\vec{c}}

\newcommand{\gv}{\vec{g}}
\newcommand{\hv}{\vec{h}}

\newcommand{\lv}{\vec{l}}

\newcommand{\nv}{\vec{n}}

\newcommand{\pv}{\vec{p}}

\newcommand{\uv}{\vec{u}}

\newcommand{\wv}{\vec{w}}
\newcommand{\xv}{\vec{x}}
\newcommand{\yv}{\vec{y}}

\newcommand{\RR}{\mathbb{R}}

%% file: acronyms.tex
\begin{acronym}
 \acro{CSI}{channel state information}
 \acro{UE}{user equipment}
 \acro{UL}{uplink}
 \acro{BS}{basestation}
 \acro{TDD}{time division duplex}
 \acro{FDD}{frequency division duplex}
 \acro{ECC}{error-correcting code}
 \acro{MLD}{maximum likelihood decoding}
 \acro{HDD}{hard decision decoding}
 \acro{IF}{intermediate frequency}
 \acro{RF}{radio frequency}
 \acro{SDD}{soft decision decoding}
 \acro{NND}{neural network decoding}
 \acro{CNN}{convolutional neural network}
 \acro{ML}{maximum likelihood}
 \acro{GPU}{graphical processing unit}
 \acro{BP}{belief propagation}
 \acro{LTE}{Long Term Evolution}
 \acro{BER}{bit error rate}
 \acro{DER}{detection error rate}
 \acro{SNR}{signal-to-noise-ratio}
 \acro{ReLU}{rectified linear unit}
 \acro{BPSK}{binary phase shift keying}
 \acro{QPSK}{quadrature phase shift keying}
 \acro{AWGN}{additive white Gaussian noise}
 \acro{MSE}{mean squared error}
 \acro{LLR}{log-likelihood ratio}
 \acro{MAP}{maximum a posteriori}
 \acro{NVE}{normalized validation error}
 \acro{BCE}{binary cross-entropy}
 \acro{CE}{cross-entropy}
 \acro{BLER}{block error rate}
 \acro{SQR}{signal-to-quantisation-noise-ratio}
 \acro{MIMO}{multiple-input multiple-output}
 \acro{OFDM}{orthogonal frequency division multiplex}
 \acro{RF}{radio frequency}
 \acro{LOS}{line of sight}
 \acro{NLoS}{non-line of sight}
 \acro{NMSE}{normalized mean squared error}
 \acro{CFO}{carrier frequency offset}
 \acro{SFO}{sampling frequency offset}
 \acro{IPS}{indoor positioning system}
 \acro{TRIPS}{time-reversal IPS}
 \acro{RSSI}{received signal strength indicator}
 \acro{MIMO}{multiple-input multiple-output}
 \acro{ENoB}{effective number of bits}
 \acro{AGC}{automated gain control}
 \acro{ADC}{analog to digital converter}
 \acro{ADCs}{analog to digital converters}
 \acro{FB}{front bandpass}
 \acro{FPGA}{field programmable gate array}
 \acro{JSDM}{Joint Spatial Division and Multiplexing}
 \acro{NN}{neural network}
 \acro{IF}{intermediate frequency}
 \acro{LoS}{line-of-sight}
 \acro{NLoS}{non-line-of-sight}
 \acro{DSP}{digital signal processing}
 \acro{AFE}{analog front end}
 \acro{SQNR}{signal-to-quantisation-noise-ratio}
 \acro{SINR}{signal-to-interference-noise-ratio}
 \acro{ENoB}{effective number of bits}
 \acro{AGC}{automated gain control}
 \acro{PCB}{printed circuit board}
 \acro{EVM}{error vector mangnitude}
 \acro{CDF}{cumulative distribution function}
 \acro{MRC}{maximum ratio combining}
 \acro{MRP}{maximum ratio precoding}
 \acro{MRT}{maximum ratio transmission}
 \acro{DeepL}{deep-learning}
 \acro{DL}{deep learning}
 \acro{SISO}{single-input single-output}
 \acro{SGD}{stochastic gradient descent}
 \acro{CP}{cyclic prefix}
 \acro{MISO}{Multiple Input Single Output}
 \acro{LMMSE}{linear minimum mean square error}
 \acro{ZF}{zero forcing}
 \acro{USRP}{universal software radio peripheral}
 \acro{RNN}{recurrent neural network}
 \acro{GRU}{gated recurrent unit}
 \acro{LSTM}{long short-term memory}
 \acro{NTM}{neural turing machine}
 \acro{DNC}{differentiable neural computer}
 \acro{TCN}{temporal convolutional network}
 \acro{FCL}{fully connected layer}
 \acro{MANN}{memory augmented neural network}
 \acro{RNN}{recurrent neural network}
 \acro{DNN}{dense neural network}
 \acro{FIR}{finite impulse response}
 \acro{BPTT}{back-propagation through time}
 \acro{GAN}{generative adversarial network}
 \acro{ELU}{exponential linear unit}
 \acro{tanh}{hyperbolic tangent}
 \acro{BICM}{bit-interleaved coded modulation}
 \acro{OTA}{over-the-air}
 \acro{IM}{intensity modulation}
 \acro{DD}{direct detection}
 \acro{RL}{reinforcement learning}
 \acro{SDR}{software-defined radio}
 \acro{WGAN}{Wasserstein generative adversarial network}
 \acro{BMD}{bit-metric decoding}
 \acro{BMI}{bit-wise mutual information}
 \acro{LDPC}{low-density parity-check}
 \acro{IDD}{iterative demapping and decoding}
 \acro{IEDD}{iterative equalization, demapping and decoding}
 \acro{JSD}{Jensen-Shannon divergence}
 \acro{MMSE}{minimum mean square error}
 \acro{FFT}{fast Fourier transform}
 \acro{IFFT}{inverse fast Fourier transform}
 \acro{QAM}{quadrature amplitude modulation}
 \acro{EMD}{earth mover's distance}
 \acro{TDL}{tapped delay line}
 \acro{KL}{Kullback-Leibler}
 \acro{PRACH}{physical random access channel}
 \acro{URLLC}{ultra-reliable low-latency communication}
 \acro{ANOMA}{asynchronous non-orthogonal multiple access}
 \acro{FEC}{forward error correction}
 \acro{NOMA}{non-orthogonal medium access}
 \acro{MTC}{machine-type communications}
 \acro{mMTC}{massive machine-type communications}
 \acro{MCS}{modulation and coding scheme}
 \acro{PAPR}{peak-to-average power ratio}
 \acro{MAC}{medium access control}
 \acro{STO}{sampling time offset}
 \acro{STE}{straight-through estimator}
 \acro{PHY}{physical}
 \acro{CCE}{categorical cross-entropy}
 \acro{IoT}{Internet of Things}
 \acro{CCDF}{complementary cumulative distribution function}
 \acro{CRC}{cyclic redundancy check}
 \acro{ACLR}{adjacent channel leakage ratio}
 \acro{MD}{missed detection}
 \acro{FA}{false alarm}
 \acro{FAR}{false alarm rate}

\end{acronym}

%% file: corporateColours.tex
\definecolor{mittelblau}{RGB}{0, 126, 198}
\definecolor{violettblau}{cmyk}{0.9, 0.6, 0, 0}
\definecolor{rot}{RGB}{238, 28 35}
\definecolor{apfelgruen}{RGB}{140, 198, 62}
\definecolor{gelb}{RGB}{1, 221, 0}
\definecolor{orange}{RGB}{244, 111, 33}
\definecolor{pink}{RGB}{237, 0, 140}
\definecolor{lila}{RGB}{128, 10, 145}
\definecolor{hellgrau}{RGB}{224, 224, 224}
\definecolor{mittelgrau}{RGB}{128, 128, 128}
\definecolor{dunkelgrau}{RGB}{80,80,80}
\definecolor{anthrazit}{RGB}{19, 31, 31}

%% file: plotcyclelists.tex
\pgfplotscreateplotcyclelist{corporate colours markers}{%
rot, every mark/.append style={fill=.!80!rot},mark=*\\%
mittelblau, every mark/.append style={fill=.!80!mittelblau},mark=square*\\%
apfelgruen, every mark/.append style={fill=.!80!apfelgruen},mark=triangle*\\%
orange, mark=star\\%
pink, every mark/.append style={fill=.!80!pink},mark=diamond*\\%
violettblau, every mark/.append style={fill=.!80!violettblau},mark=otimes*\\%
lila, mark=|\\%
gelb, every mark/.append style={fill=.!80!gelb},mark=pentagon*\\%
hellgrau, mark=text,text mark=p\\%
anthrazit, mark=text,text mark=a\\%
}

\pgfplotscreateplotcyclelist{corporate colours markers double}{%
rot, every mark/.append style={fill=.!80!rot},mark=*\\%
rot, dashed, every mark/.append style={fill=.!80!rot,solid},mark=*\\%
mittelblau, every mark/.append style={fill=.!80!mittelblau},mark=square*\\%
mittelblau, dashed, every mark/.append style={fill=.!80!mittelblau,solid},mark=square*\\%
apfelgruen, every mark/.append style={fill=.!80!apfelgruen},mark=triangle*\\%
apfelgruen, dashed, every mark/.append style={fill=.!80!apfelgruen,solid},mark=triangle*\\%
orange, mark=star\\%
orange, dashed, mark=star\\%
pink, every mark/.append style={fill=.!80!pink},mark=diamond*\\%
pink, dashed, every mark/.append style={fill=.!80!pink,solid},mark=diamond*\\%
violettblau, every mark/.append style={fill=.!80!violettblau},mark=otimes*\\%
violettblau, dashed, every mark/.append style={fill=.!80!violettblau,solid},mark=otimes*\\%
lila, mark=|\\%
lila, dashed, mark=|\\%
gelb, every mark/.append style={fill=.!80!gelb},mark=pentagon*\\%
gelb, dashed, every mark/.append style={fill=.!80!gelb,solid},mark=pentagon*\\%
}

\pgfplotsset{
  colormap/magma/.style={%
    /pgfplots/colormap={magma}{%
      rgb=(0.001462, 0.000466, 0.013866)
      rgb=(0.035520, 0.028397, 0.125209)
      rgb=(0.102815, 0.063010, 0.257854)
      rgb=(0.191460, 0.064818, 0.396152)
      rgb=(0.291366, 0.064553, 0.475462)
      rgb=(0.384299, 0.097855, 0.501002)
      rgb=(0.475780, 0.134577, 0.507921)
      rgb=(0.569172, 0.167454, 0.504105)
      rgb=(0.664915, 0.198075, 0.488836)
      rgb=(0.761077, 0.231214, 0.460162)
      rgb=(0.852126, 0.276106, 0.418573)
      rgb=(0.925937, 0.346844, 0.374959)
      rgb=(0.969680, 0.446936, 0.360311)
      rgb=(0.989363, 0.557873, 0.391671)
      rgb=(0.996580, 0.668256, 0.456192)
      rgb=(0.996727, 0.776795, 0.541039)
      rgb=(0.992440, 0.884330, 0.640099)
      rgb=(0.987053, 0.991438, 0.749504)
    },
  },
  colormap/inferno/.style={%
    /pgfplots/colormap={inferno}{%
      rgb=(0.001462, 0.000466, 0.013866)
      rgb=(0.037668, 0.025921, 0.132232)
      rgb=(0.116656, 0.047574, 0.272321)
      rgb=(0.217949, 0.036615, 0.383522)
      rgb=(0.316282, 0.053490, 0.425116)
      rgb=(0.410113, 0.087896, 0.433098)
      rgb=(0.503493, 0.121575, 0.423356)
      rgb=(0.596940, 0.154848, 0.398125)
      rgb=(0.688653, 0.192239, 0.357603)
      rgb=(0.775059, 0.239667, 0.303526)
      rgb=(0.851384, 0.302260, 0.239636)
      rgb=(0.912966, 0.381636, 0.169755)
      rgb=(0.956852, 0.475356, 0.094695)
      rgb=(0.981895, 0.579392, 0.026250)
      rgb=(0.987464, 0.690366, 0.079990)
      rgb=(0.973088, 0.805409, 0.216877)
      rgb=(0.947594, 0.917399, 0.410665)
      rgb=(0.988362, 0.998364, 0.644924)
    },
  },
  colormap/plasma/.style={%
    /pgfplots/colormap={plasma}{%
      rgb=(0.050383, 0.029803, 0.527975)
      rgb=(0.186213, 0.018803, 0.587228)
      rgb=(0.287076, 0.010855, 0.627295)
      rgb=(0.381047, 0.001814, 0.653068)
      rgb=(0.471457, 0.005678, 0.659897)
      rgb=(0.557243, 0.047331, 0.643443)
      rgb=(0.636008, 0.112092, 0.605205)
      rgb=(0.706178, 0.178437, 0.553657)
      rgb=(0.768090, 0.244817, 0.498465)
      rgb=(0.823132, 0.311261, 0.444806)
      rgb=(0.872303, 0.378774, 0.393355)
      rgb=(0.915471, 0.448807, 0.342890)
      rgb=(0.951344, 0.522850, 0.292275)
      rgb=(0.977856, 0.602051, 0.241387)
      rgb=(0.992541, 0.687030, 0.192170)
      rgb=(0.992505, 0.777967, 0.152855)
      rgb=(0.974443, 0.874622, 0.144061)
      rgb=(0.940015, 0.975158, 0.131326)
    },
  },
}

%% file: tikz/bl_message_5.tikz
\tikzfading[name=fadelr,left color=blue!0, right color=red!100]
\tikzfading[name=faderl,left color=blue!100, right color=red!0]

\tikzfading[name=fadeud,top color=blue!0, bottom color=red!100]
\tikzfading[name=fadedu,top color=blue!100, bottom color=red!0]

\includegraphics{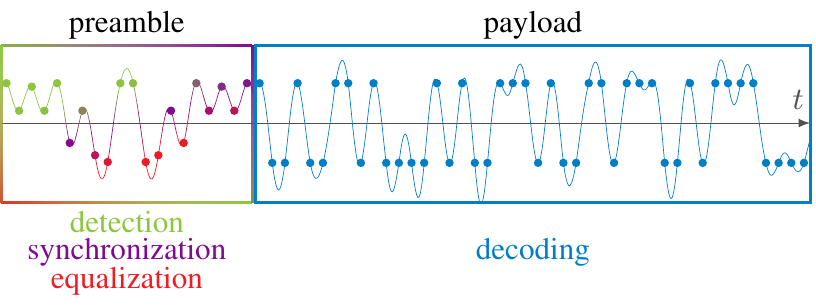}

%% file: tikz/phyae_message_7.tikz
\tikzfading[name=fadelr,left color=blue!0, right color=red!100]
\tikzfading[name=faderl,left color=blue!100, right color=red!0]

\tikzfading[name=fadeud,top color=blue!0, bottom color=red!100]
\tikzfading[name=fadedu,top color=blue!100, bottom color=red!0]

\includegraphics{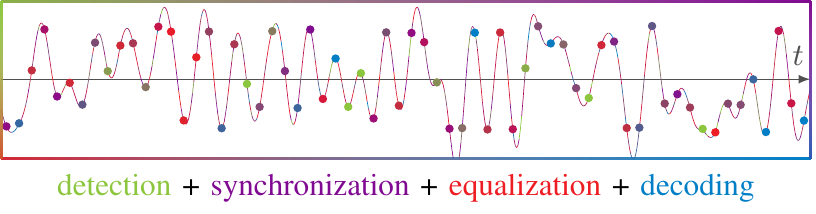}

%% file: tikz/system_model.tex
\includegraphics{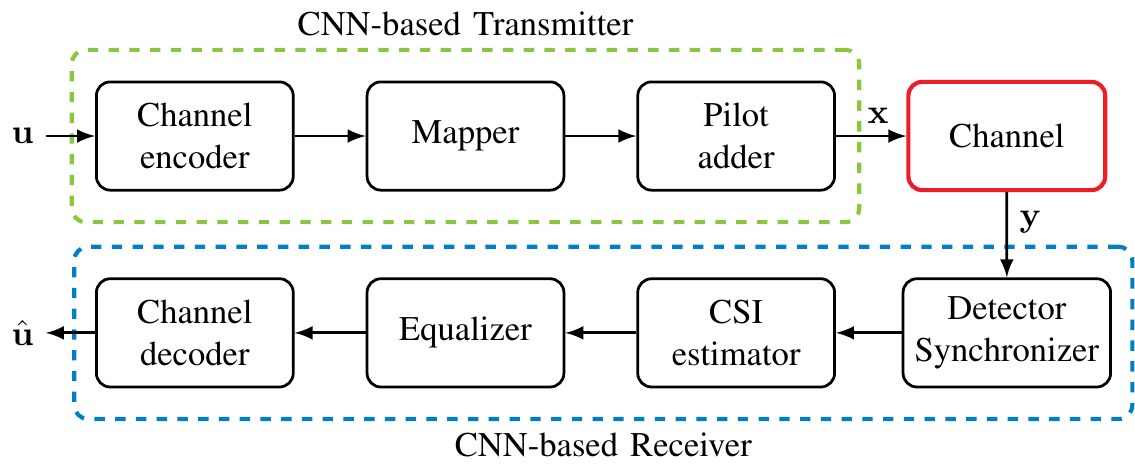}

%% file: tikz/channel_model_tikz.tex
\includegraphics{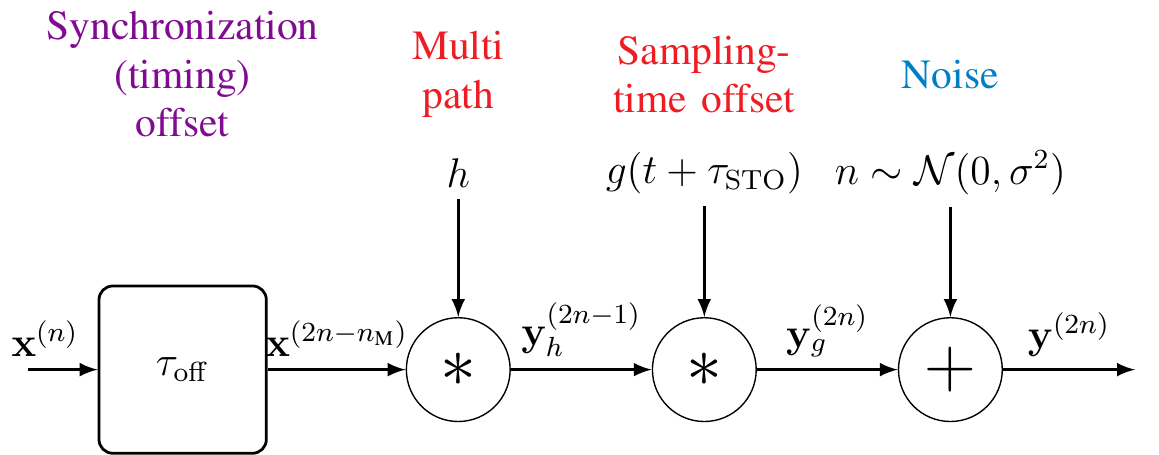}

%% file: tikz/baseline_RX_system_tikz.tex
\includegraphics{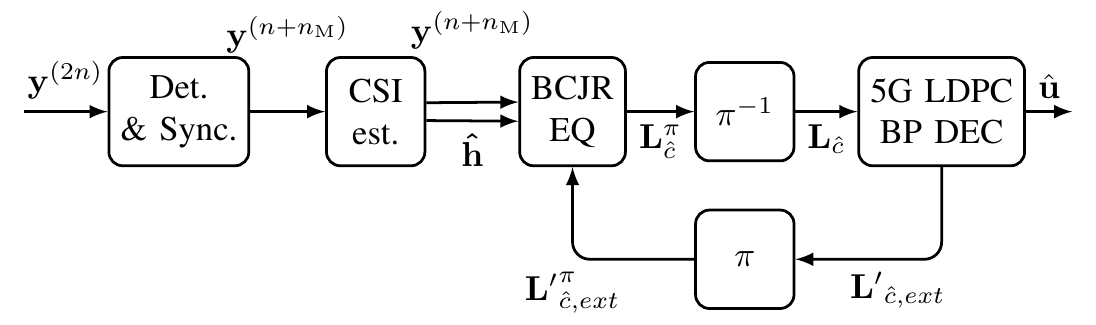}

%% file: tikz/serial_encoder.tex
\includegraphics{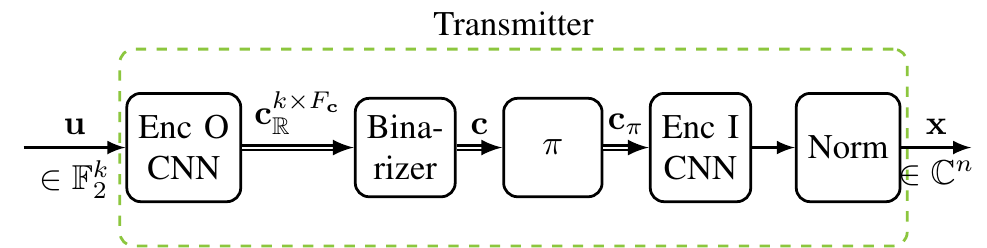}

%% file: tikz/serial_decoder.tex
\includegraphics{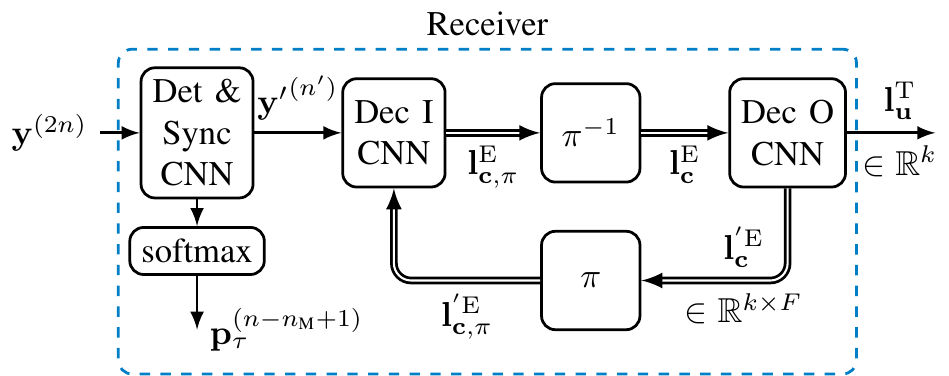}

%% file: tikz/phyae_detector.tikz
\tikzfading[name=fadelr,left color=blue!0, right color=red!100]
\tikzfading[name=faderl,left color=blue!100, right color=red!0]

\tikzfading[name=fadeud,top color=blue!0, bottom color=red!100]
\tikzfading[name=fadedu,top color=blue!100, bottom color=red!0]

\includegraphics{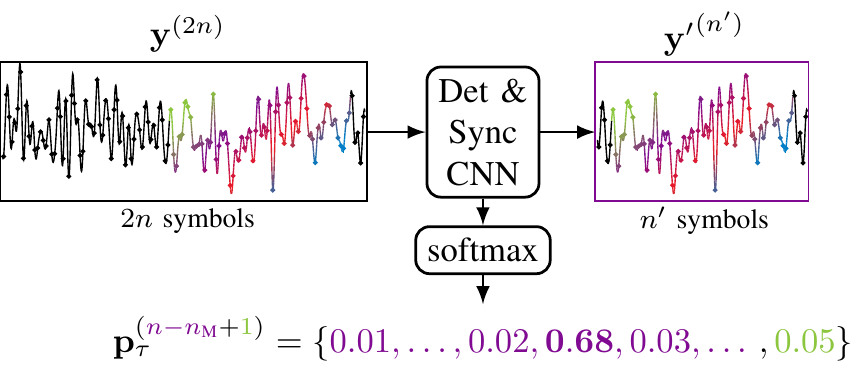}

%% file: tikz/hist_unconstrained.tikz
\includegraphics{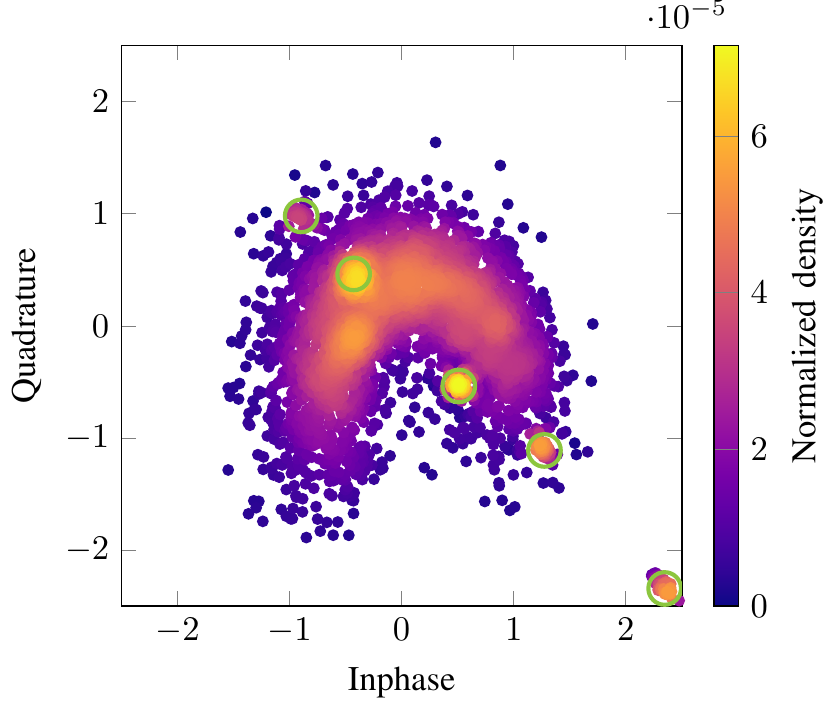}

%% file: tikz/learned_const_energy.tikz
\includegraphics{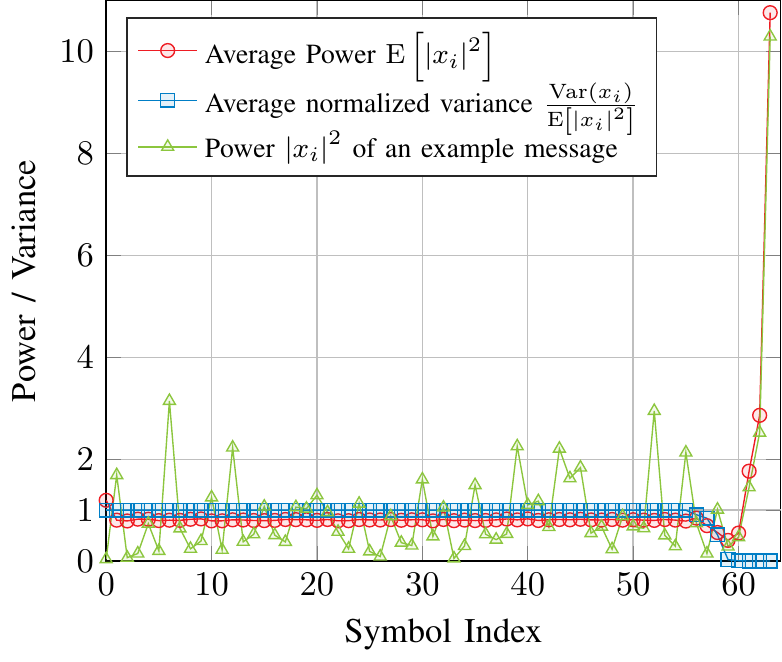}

%% file: tikz/hist_constrained.tikz
\includegraphics{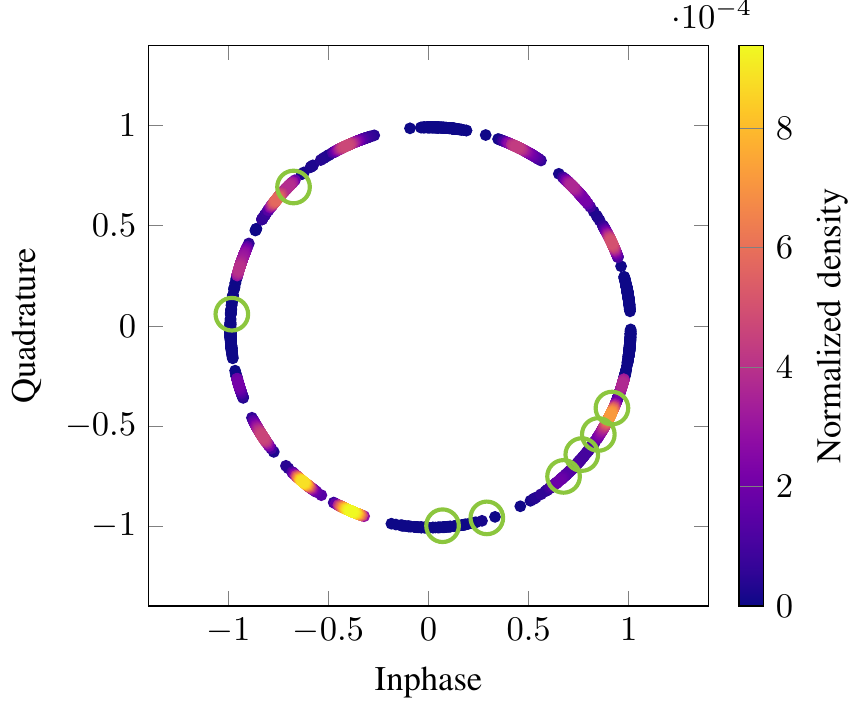}

%% file: tikz/papr_ccdf.tikz
\includegraphics{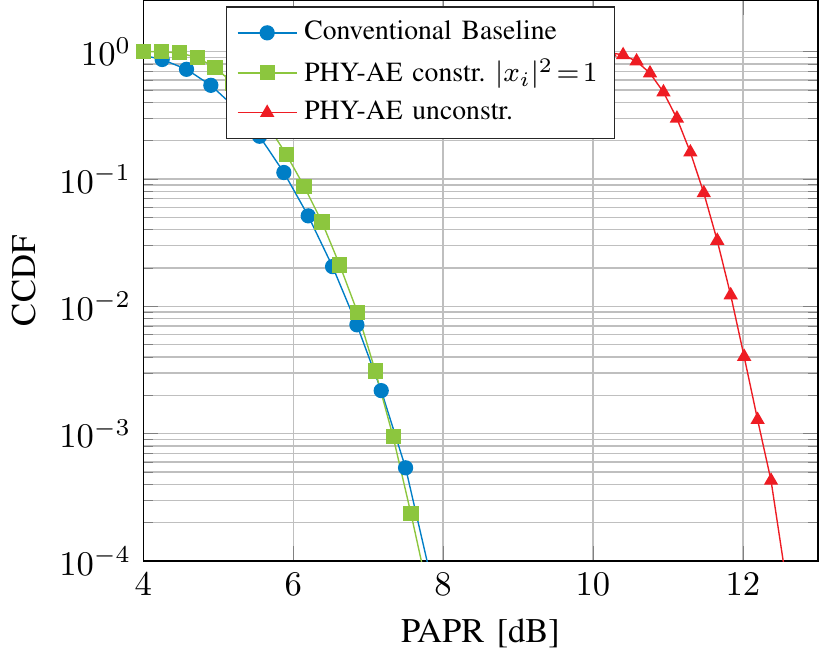}

%% file: tikz/der_prach.tikz
\includegraphics{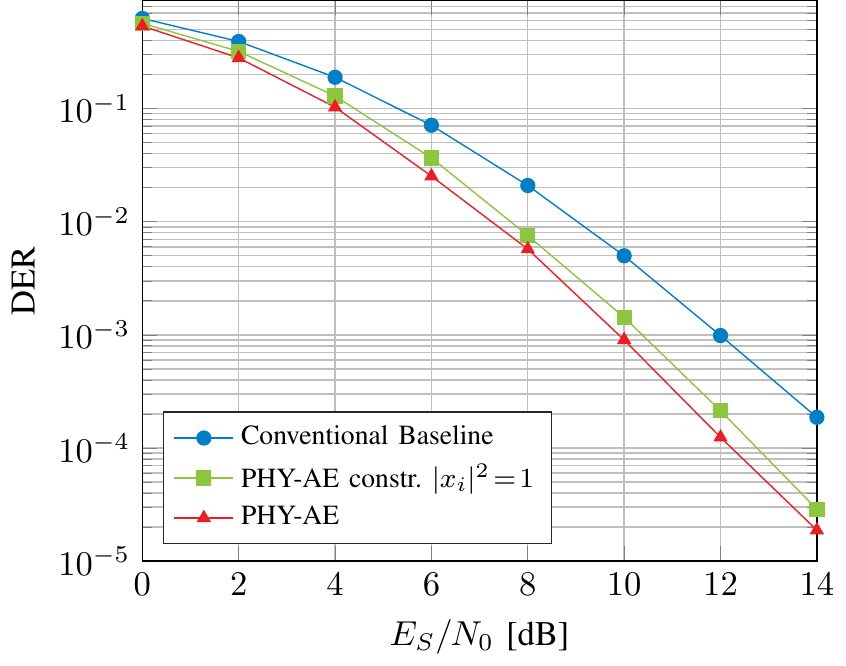}

%% file: tikz/ber_prach_toff_sim.tikz
\includegraphics{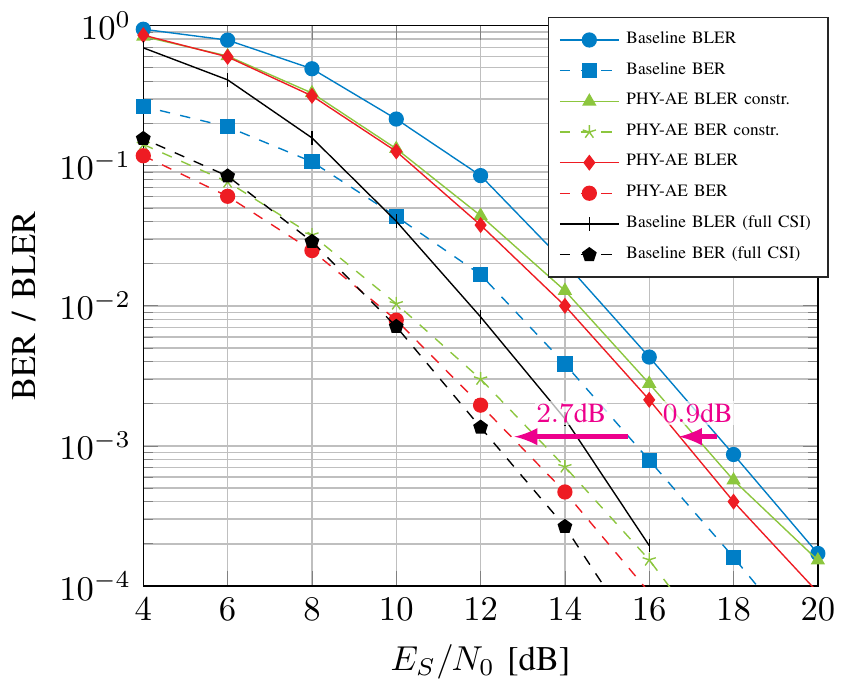}

%% file: tikz/ber_prach_sim_h.tikz
\includegraphics{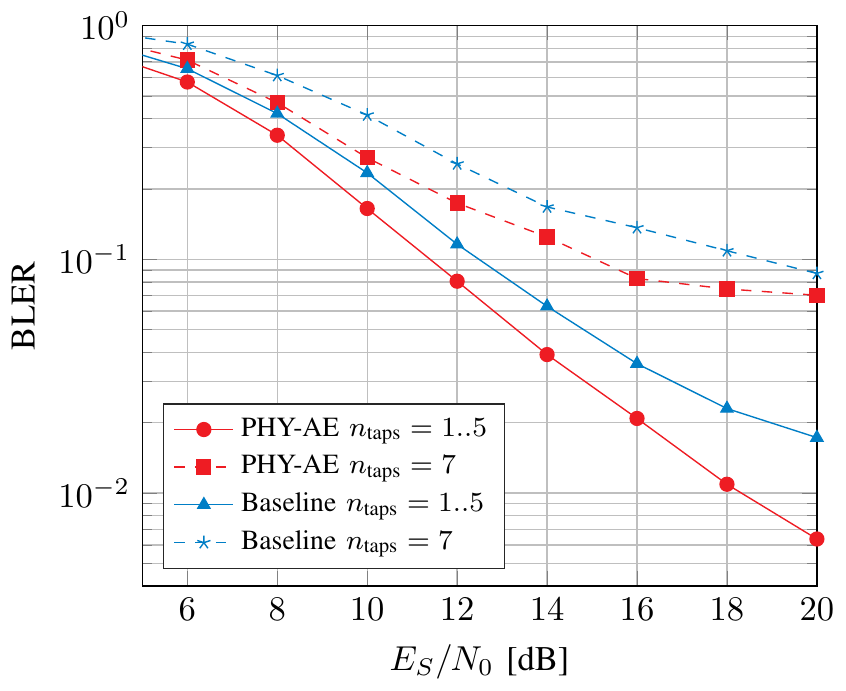}

%% file: tikz/ber_length_sweep.tikz
\includegraphics{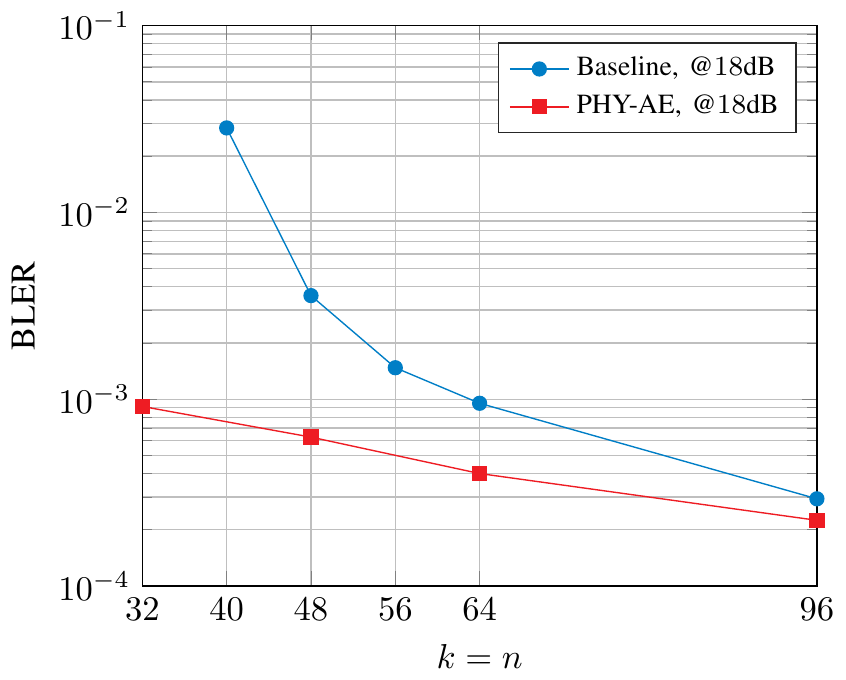}

%% file: tikz/channel_h.tikz
\includegraphics{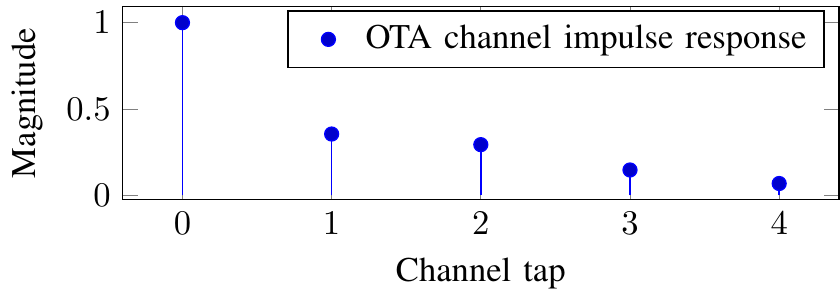}

%% file: tikz/der_prach_real_ch.tikz
\includegraphics{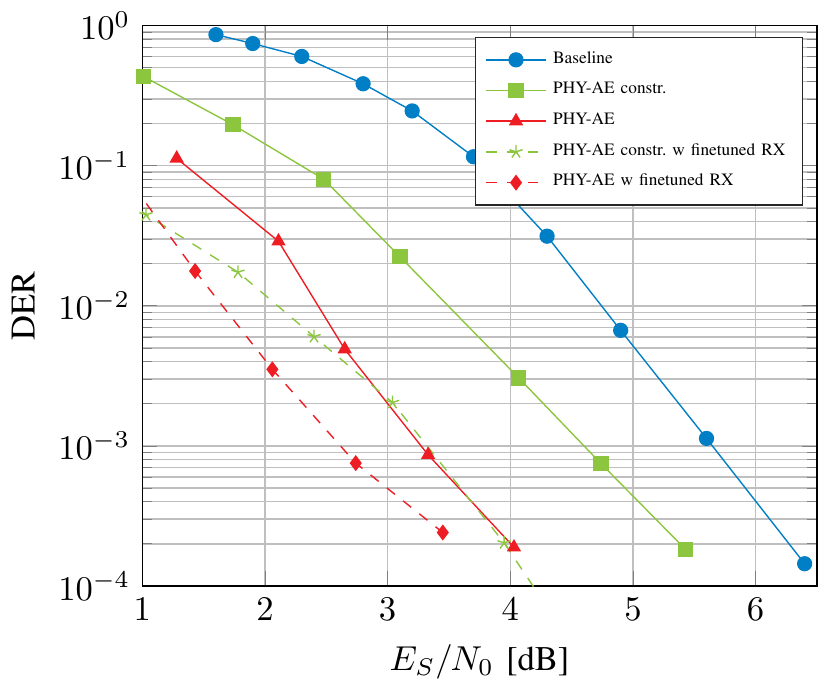}

%% file: tikz/ber_prach_real.tikz
\includegraphics{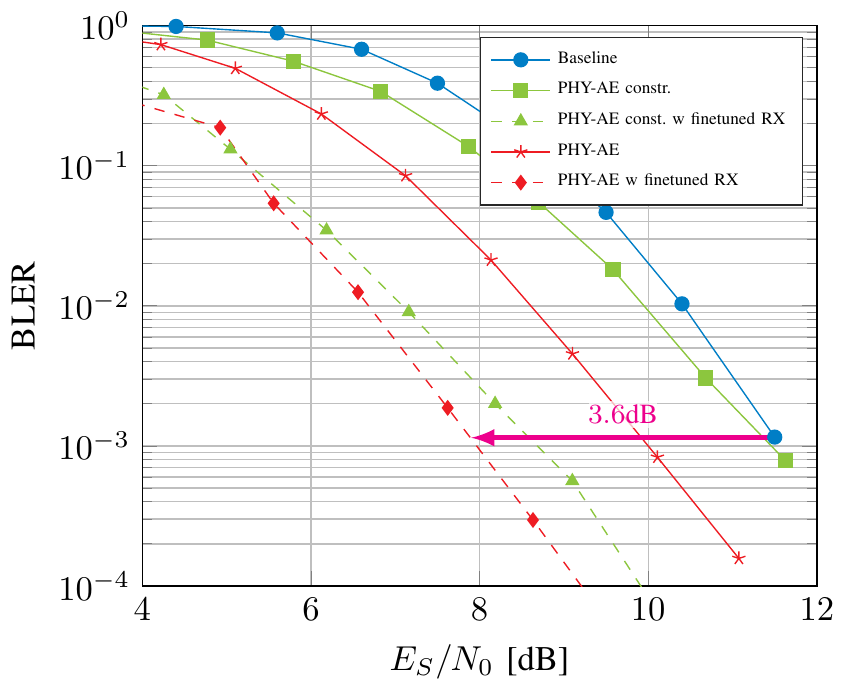}

%% file: appendix/cnn_block.tikz
\includegraphics{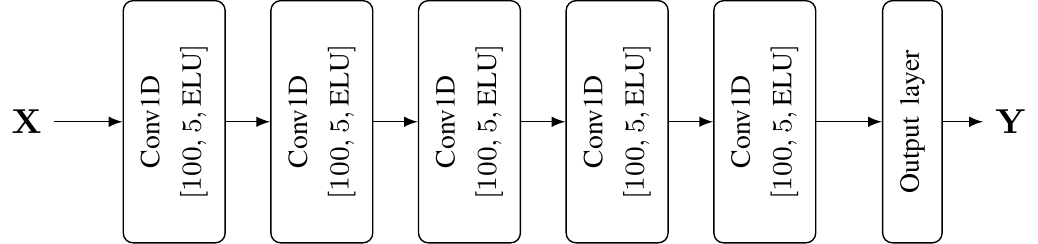}

%% file: appendix/generator_structure.tikz
\includegraphics{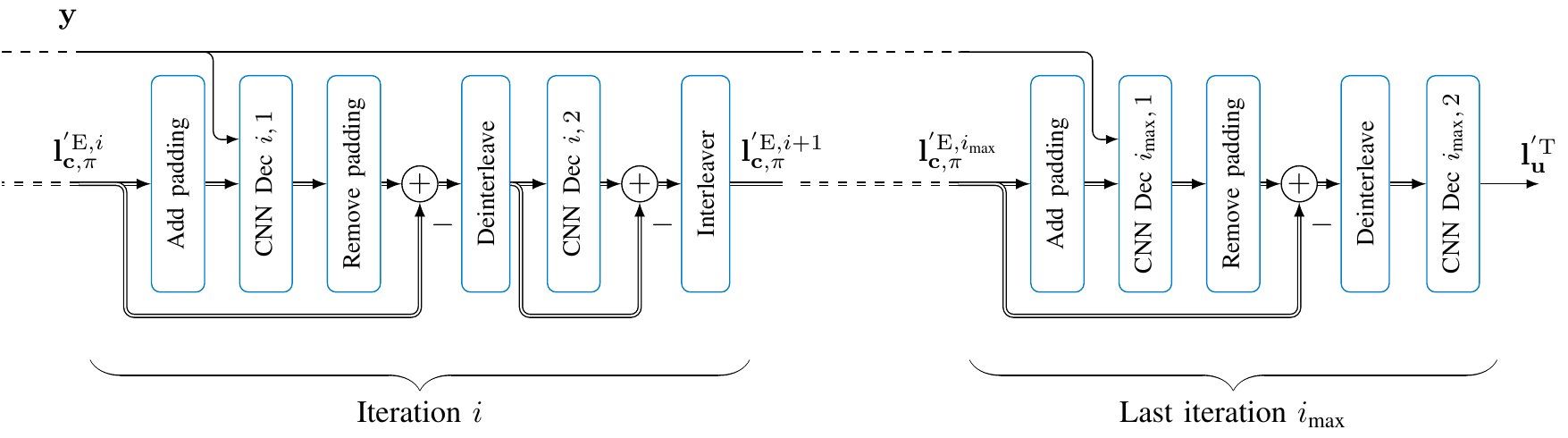}